%Paper: hep-th/9510003
%From: imbimbo@surya23.cern.ch (Camillo Imbimbo)
%Date: Mon, 2 Oct 1995 16:04:49 --100

\input harvmac
\overfullrule=0pt
%Definitions
\def\np{Nucl. Phys.}
\def\pl{Phys. Lett.}
\def\prl{ Phys. Rev. Lett.}
\def\cmp{ Commun. Math. Phys.}

%Operators

\def\Lv{L_{\hat v}}

\def\Ia#1{I_{a_{#1}}}

\def\Lv#1{{\cal L}_{{\hat v}_{#1}}}
\def\lva{{\cal L}_{{\hat v}_a}}
\def\lvb{{\cal L}_{{\hat v}_b}}

\def\leao{(\lva \het{0})^{\mu\nu}}
\def\lebo{(\lvb \het{0})^{\mu\nu}}
\def\vIa#1{\check{I}_{a_{#1}}}

% vector fields

\def\va{{\hat v}_a}
\def\vb{{\hat v}_b}
\def\vac{{\tilde v}_a}
\def\vbc{{\tilde v}_b}

% fields
\def\abeta{\beta_{\mu \nu}}
\def\gmunu{g_{\mu\nu}}
\def\hatg{{\hat g}}
\def\ghat{{\hat g}^{\mu\nu}}
\def\ino{{\psi}_{\mu\nu}}
\def\inoh{{\hat \psi}^{\mu\nu}}
\def\hino{{\hat\psi}}

\def\lc{{\cal L}_c}

\def\lgamma{{\cal L}_{\gamma}}
\def\ab{b_{\mu \nu}}
\def\ci{{\cal C}^i}
\def\Phio{\Phi_0}
\def\Phia{\Phi_a}
\def\Phib{\Phi_b}

%background metric

\def\ehat{{\hat\eta}^{\mu\nu}}
\def\ehatd{{\hat{\eta}}_{\mu\nu}}
\def\emunu{\eta_{\mu\nu}}
\def\emunua{\eta^{\mu\nu}_a}
\def\emunub{\eta^{\mu\nu}_b}
\def\het#1{{\hat\eta}_{#1}}
\def\ehatb{{\hat\eta}^{\mu\nu}_b}
\def\ehata{{\hat\eta}^{\mu\nu}_a}
\def\ehato{{\hat\eta}^{\mu\nu}_0}
\def\dpeo{d_p\ehato}
\def\If#1{{\lambda}_{#1}}
\def\bIf#1{{\bar\lambda}_{#1}}
\def\Map#1{\Phi_m^{(#1)}}
\def\bel#1{\mu_{#1}}
\def\bbel#1{{\bar\mu}_{#1}}

% Cell decomposition

\def\mgn{{\cal M}_{g,n}}

\def\mfour{{\cal M}_{0,4}}
\def\Ua{{\cal U}_a}
\def\Ub{{\cal U}_b}
\def\UU#1{{\cal U}_{#1}}
\def\Uab{{\cal U}_{ab}}
\def\Uq{{\cal U}_{a_0a_1\ldots a_q}}
\def\Cel#1{{\cal C}_{#1}}
\def\Cq{{\cal C}_{a_0a_1\ldots a_q}}

% forms

\def\Zf#1{Z^{(2N-{#1})}_{#1}}
\def\Zqplus{Z^{(2N-q-1)}_{q+1}}
\def\ZN{Z^{(N)}_N}
\def\SN{S^{(N)}_{N-1}}
\def\Zmp{Z(m^i;p^i)}
\def\Zmpa{Z_a (m^i;p^i)}
\def\Zmpb{Z_b (m^i;p^i)}
\def\Sab{S_{ab} (m^i;p^i)}
\def\Zglobal{Z_0^{global}}
\def\Zga{(\Zglobal)_a}
\def\Zg{Z^{global}}
\def\xa{x_a}
\def\xb{x_b}
\def\x0{x_0}

\def\Li{{\cal L}_i}
\def\Zm{Z_m}
\def\Zmbar{{\bar Z}_m}
\def\za{z_a}
\def\zabar{{\bar z}_a}

%matrices

\def\Ma{{\rm M}_a}
\def\Mai{{\rm M}_a^{-1}}
\def\Mbi{{\rm M}_b^{-1}}

\def\Mb{{\rm M}_b}

% references

\def\LPW{J. Labastida, M. Pernici and E. Witten, \np\ {\bf B 310} (1988)
611.}
\def\MoSo{D. Montano and J. Sonnenschein, \np\ {\bf B 313} (1989) 258.}
\def\MYPE{R. Myers and V. Periwal, \np\ {\bf B 333} (1990) 536.}
\def\HV{H. Verlinde, Utrecht preprint THU-87/26, 1987, unpublished.}
\def\VV{E. Verlinde and H. Verlinde, \np\ {\bf B 348} (1991) 457.}

\def\MYE{R. Myers, \np\ {\bf B 343} (1990) 705.}
\def\BMS{R. Brooks, D. Montano and J. Sonnenschein, \pl\ {\bf B 214}
(1988) 91.}
\def\MS{D. Montano and J. Sonnenschein, \np\ {\bf B 324} (1989) 348.}
\def\BS{L. Baulieu and I.M. Singer, \cmp\ {\bf 135} (1991) 253.}
\def\BCOV{M. Bershadsky, S. Cecotti, H. Ooguri and C. Vafa,
hep-th/9302103, \np\ {\bf B 405} (1993) 279.}
\def\BCI{C.M. Becchi, R. Collina and C. Imbimbo, hep-th/9311097,
\pl\ {\bf B 322} (1994) 79.}
\def\TORINO{C.M. Becchi, R. Collina and C. Imbimbo, hep-th/9406096,
CERN and Genoa preprints CERN-TH 7302/94, GEF-Th 6/1994,
{\it Symmetry and Simplicity in Theoretical Physicis},
Proceedings of the Symposium for the 65-th Birthday of Sergio Fubini,
Turin, 1994 (World Scientific, Singapore, 1994).}
\def\TRIESTE{ C. Imbimbo, \np\ (Proc. Suppl.) {\bf B 41} (1995) 302.}
\def\BOTT{R. Bott and L.W. Tu, {\it Differential Forms in Algebraic
Topology} (Springer-Verlag, New York, 1982).}
\def\NELSON{P. Nelson, \prl\ {\bf 62} (1989) 993.}
\def\DN{J. Distler and P. Nelson, \cmp\ {\bf 138} (1991) 273.}
\def\MORITA{S. Morita, Invent. Math. {\bf 90} (1987) 551.}
\def\MUMFORD{D. Mumford, ``Towards an enumerative geometry of the
moduli space of curves'', in {\it Arithmetic and Geometry},
Michael Artin and John Tate, eds. (Birkh\"auser, Basle, 1983).}
\def\MILLER{E. Miller, J. Diff. Geom. {\bf 24} (1986) 1.}
\def\KONSEVITCH{M. Konsevitch, \cmp\ {\bf 147} (1992) 1.}

\def\WITTOP{ E. Witten, \np\ {\bf B 340} (1990) 281.}
\def\WITDI{R. Dijkgraaf and E. Witten, \np\ {\bf B 342} (1990) 486.}

\def\BB{L. Baulieu and M. Bellon, \pl\ {\bf B 202} (1988) 67.}
\def\BZ{O. Bergman and B. Zwiebach, hep-th/9411047, \np\ {\bf B 441}
(1995) 76.}
\def\PO{J. Polchinsky, \np\ {\bf B 307} (1988) 61.}
\def\LI{K. Li, \np\ {\bf B 354} (1990) 711 and 725.}
\def\KMINUS{A. Losev, hep-th/9211089, Theor. Math. Phys {\bf 95} (1993)
595.\semi
D. Ghoshal and S. Mukhi, hep-th/9312189, \np\ {\bf B 425} (1994)
173\semi A. Hanany, Y. Oz and R. Plesser, hep-th/9401030, \np\
{\bf B 425} (1994) 150\semi
Y. Lavi, Y. Oz and J. Sonnenschein, hep-th/9406056, \np\
{\bf B 431} (1994) 223\semi
D. Ghoshal, C. Imbimbo and S. Mukhi, hep-th/9410034, \np\ {\bf B 440}
(1995) 355.}
\def\CMR{S. Cordes, G. Moore and S. Ramgoolam, hep-th/9402107, Yale
preprint YCTP-P23-93, 1994.}

% text

{\nopagenumbers
\null
\vskip -2truecm
\abstractfont\hsize=\hstitle\rightline{
\vtop{
\hbox{hep-th/9510003}
\hbox{CERN-TH/95-242}
\hbox{GEF-TH/95-8}}}
\pageno=0
\vskip 1truecm
\centerline{\titlefont Gribov problem, contact terms and
\v{C}ech-De Rham}
\bigskip
\centerline{\titlefont cohomology in 2D topological gravity}
\bigskip
\centerline{Carlo M. Becchi\foot{E-mail: becchi@genova.infn.it}}
\vskip 4pt
\centerline{\it Dipartimento di Fisica dell' Universit\'a di Genova}

\centerline{\it Via Dodecaneso 33, I-16146, Genova, Italy}
\vskip 7pt
\centerline{Camillo Imbimbo\foot{E-mail: imbimbo@vxcern.cern.ch}}
\vskip 4pt
\centerline{{\it Theory Division, CERN, CH-1211 Geneva 23,
Switzerland}\foot{On leave from INFN, Sezione di Genova, Genoa, Italy.}}
%\vskip 7pt
\ \medskip
\centerline{ABSTRACT}

We point out that averages of equivariant observables of 2D
topological gravity are not globally defined forms on moduli space,
when one uses the functional measure corresponding to the formulation
of the theory as a 2D superconformal model. This is shown to be a
consequence of the existence of the Gribov horizon {\it and} of the
dependence of the observables on derivatives of the super-ghost field.
By requiring the absence of global BRS anomalies, it is nevertheless
possible to associate global forms to correlators
of observables by resorting to the \v{C}ech-De Rham notion of form
cohomology. To this end, we derive and solve the ``descent'' of local
Ward identities which characterize the functional measure. We obtain
in this way an explicit expression for the \v{C}ech-De Rham cocycles
corresponding to arbitrary correlators of observables.  This provides
the way to compute and understand contact terms in string theory from
first principles.
\ \vfill
\ifx\answ\bigans
\leftline{CERN-TH/95-242}
\leftline{GEF-TH/95-8}
\leftline{September 1995}
\else\leftline{September1995}\fi
\eject}
\ftno=0

\newsec{Introduction}

In the modern formulation of closed string theory,
$n$-point amplitudes of order $g$ are integrals of closed
top-forms over $\mgn$ -- the moduli space of Riemann surfaces of genus
$g$ and $n$ marked points. These top-forms are correlators of operators
that belong in the BRS cohomology of the underlying
2D quantum field theory. $\mgn$ is non-compact: its points at
``infinity'' correspond to degenerate surfaces with
nodes. ``Contact'' terms are usually referred to as
the contributions to the string amplitudes coming from those singular
geometries.

The familiar example illustrating the role of contact terms
in string theory is the dilaton equation, a low-energy recursion
relation expected to be valid in any closed string theory\ref\po{\PO}:
\eqn\dilaton{
\int_{{\cal M}_{g,n+1}}\Bigl\langle \sigma_1^{(0)}(x) \prod_{i=1}^n
O_i^{(0)}(x_i)\Bigr\rangle
= (2g-2+n)\int_{\mgn}\Bigl \langle \prod_{i=1}^n O_i^{(0)}(x_i)\Bigr
\rangle.}
In Eq. \dilaton\ the $O_i^{(0)}(x_i)$ are generic observables with
values in the $0$-forms on the world-sheet.
The (zero-momentum) dilaton operator $\sigma_1^{(0)}(x)$
is the element of the cohomology of the BRS operator $s$ which is
obtained from the world-sheet Euler $2$-form $\sigma_1^{(2)} = R^{(2)}$
via the so-called ``descent'' equations:
\eqn\descent{
s \sigma_1^{(2)} = d\sigma_1^{(1)}, \quad s\,\sigma_1^{(1)} =
d \sigma_1^{(0)},\quad s \sigma_1^{(0)} =0.}

Equation \dilaton\ is usually understood \po,\ref\nelson{\NELSON},
\ref\dn{\DN}\ by interpreting the $2g-2$ in
the r.h.s. as the ``bulk'' part of the amplitude. The $n$
term is instead seen to arise from contacts between $\sigma_1^{(0)}(x)$
and the operators $O^{(0)}_i (x_i)$ when the modulus associated with the
point $x$ approaches the punctures $x_i$.

Topological string theories describing non-critical strings with
$c\leq 1$ have an infinite number of dilaton-like observables -- the
gravitational descendants -- whose correlators obey recursion
relations \ref\wittop{\WITTOP},\ref\witdi{\WITDI},\ref\vv{\VV}\
which generalize the dilaton equation \dilaton.
These recursion relations can be
interpreted, in much the same way as the dilaton equation, by
decomposing physical amplitudes into ``bulk'' (usually
simple to evaluate) and  ``contact'' parts (typically much more
difficult to derive from first
principles)\vv--\nref\li{\LI}\ref\kminus{\KMINUS}.
Contact terms have been shown to encode much of the
dynamics of topological string models describing large-$N$ 2D
Yang-Mills theory as well \ref\cmr{\CMR}. Finally, the study of
contact terms in models of topological gravity coupled to $N=2$
topological models has also led to the discovery of the
```holomorphic'' anomaly \ref\bcov{\BCOV}.

It seems fair to say that the world-sheet, field theoretical
understanding of contact terms in (topological) string theories has so
far remained heuristic. In concrete situations, contact terms have
been computed mainly through consistency requirements and/or
comparison with known a priori results. It would seem reasonable to
expect that contact terms be determined unambiguously by
gauge-invariance considerations. Instead, even the derivation of the
dilaton equation (the best understood among the recursion relations of
string theory) as presented in \dn, involves certain arbitrary choices
-- emphasized, for example, in \ref\bz{\BZ}.

In the present article, we consider the issue of contact terms in
the context of 2D topological gravity. Here the situation is somewhat
paradoxical. The solution of this model presented in Ref. \vv\
starts from a Lagrangian describing a free superconformal
theory and from a set of physical observables. Averages of these
observables vanish identically in the corresponding
functional measure. The whole
non-trivial content of the theory lies in contacts terms
whose structure is derived from factorization arguments rather
than from a direct evaluation of the functional integral.

It seems  natural to wonder if (and how) the information about
such contacts is encoded in the original -- vanishing -- functional
measure,
or if extra, hidden choices, beyond that of the Lagrangian,
are necessary to determine the values of the contacts.

In this paper we show that the physical principle that fixes
contact terms ambiguities is the vanishing of {\it global} BRS anomalies.
This requirement turns out to be  non-trivial since we
will prove that averages of observables -- though generally vanishing
in the functional measure considered--
are {\it not} globally defined on $\mgn$.
The technical novelty of our analysis with respect to previous treatments
of topological gravity is formulating the quantum theory in a
general covariant background gauge. This allows us to probe the
dependence of correlators on the gauge-slice.

A familiar feature of non-Abelian gauge theories is the
existence the Gribov horizon, that is the lack of a global gauge-slice.
The Gribov horizon is the locus in orbit space where the Fadeev-Popov
measure degenerates. This would coincide, in the case of topological
gravity,
with moduli space itself, since the antighost zero modes
cause the Fadeev-Popov measure relative to the local fields to vanish
identically. In order to gauge-fix these zero modes, the action of
BRS operator needs to be extended to global quantum mechanical variables
associated to the moduli and supermoduli. Since string theorists
might be unacquainted with such
an explicitly BRS-invariant procedure to gauge-fix the antighost
zero modes, we review it in section 2.

The new Fadeev-Popov measure, including both the local fields and
the global quantum mechanical variables, is generically non-degenerate.
The Gribov horizon associated to it has codimension one in moduli space.
This implies that the functional integral defines correlators of
observables which are {\it local} closed top-forms on $\mgn$.
In order that the correlators have physical meaning, however, such
locally defined forms must be local restrictions of forms which are
globally defined. For this to be true the observables must satisfy
suitable conditions.
In closed string theory this condition is known, in the context of the
conformal gauge, as the $b_0^-$ equivariance condition \nelson.
In a covariant framework the $b_0^-$ condition is equivalent to the
request that observables be reparametrization-invariant \ref\bci{\BCI}.
Nevertheless, in the case of topological gravity,
functional averages of equivariant (i.e. reparametrization-covariant)
observables are not in general globally defined.
We will demonstrate this by deriving the Ward identities
associated to {\it finite} reparametrizations of the background
gauge. This phenomenon is originated by the dependence of the
observables on derivatives of the super-ghost field.

Even if functional averages of equivariant observables are not globally
defined, it is still possible to associate to them
globally defined forms by resorting to the \v{C}ech-De Rham
notion of form cohomology. The idea is to derive from our ``anomalous''
Ward identities a chain of descendant identities defining
a local cocycle of the \v{C}ech-De Rham complex of $\mgn$.
A well-known construction of
cohomology theory leads from this local cocycle to a globally
defined form. The integral of the globally defined form
receives contributions not only from the original local top-form
(which vanishes in the superconformal gauge),
but also from the tower of local forms
of lower degree that solve the chain of Ward identities.
The contact terms are, in this view, precisely the
contributions of the descendant local forms to the globally
defined integral. They compensate the lack
of global definition of the original top-form correlator.

The paper is organized as follows: in section 2, we review
the definition of the theory and of the observables, and
present the construction of the Lagrangian in
a reparametrization-covariant framework.
Essentially, this Lagrangian is the background-gauge
version of the Lagrangian of Ref. \vv\ which
was written in the superconformal gauge.  In section 3,
we derive and solve the Ward identities relative to finite
reparametrizations of the background metric. In section 4,
we recall the notion  of \v{C}ech-De Rham cohomology and
exhibit its relevance to the case at hand.
By solving the \v{C}ech-De Rham chain of Ward identities
we arrive at an explicit expression for the \v{C}ech-De Rham
cocycles associated to arbitrary correlators of observables.
In section 5, we show that the \v{C}ech-De Rham expression
for correlators we obtained from the local Ward identities
agrees with the algebro-geometric definition of correlators in
terms of intersection numbers of cohomology classes on moduli
space. In sections 6 and 7 we apply our general formulas
to some simple correlators to verify the agreement with
the dilaton and puncture equations of Refs. \wittop,\vv.
The goal of this exercise is to show that in this framework
the computation of contact terms does not involve ambiguities
or arbitrary choices, but is completely combinatoric.

\newsec{Lagrangian and BRS identities}

Two-dimensional topological gravity
\ref\lpw{\LPW},\ref\moso{\MoSo}\ is a
topological quantum field theory characterized by the following
nilpotent BRS transformation laws
\ref\mype{\MYPE},\ref\bms{\BMS},\vv :
\eqn\brs{\eqalign{&s\, \gmunu = \lc \gmunu + \ino \cr
&s\, \ino = \lc \ino - \lgamma \gmunu \cr}\qquad
\eqalign{&s\, c^{\mu}\, = \half \lc c^{\mu} + \gamma^{\mu} \cr
&s\, \gamma^{\mu}=\lc \gamma^{\mu},\cr}}
where $\gmunu$\ is the two-dimensional metric, $\ino$ is the gravitino
field, $c^{\mu}$ is the ghost vector field and $\gamma^{\mu}$ is
the superghost vector field; $\lc$ and $\lgamma$ denote the action of
infinitesimal diffeomorphisms with parameters $c^{\mu}$ and
$\gamma^{\mu}$ respectively.

A class of observables, local in the fields $\gmunu ,\ino ,c^{\mu}$
and $\gamma^{\mu}$, can be constructed
\mype ,\ref\mye{\MYE},\ref\ms{\MS},\ref\bs{\BS},\vv,\bci\
starting from the Euler two-form
\eqn\euler{\sigma^{(2)} = {1\over 8\pi}\sqrt{g} R \epsilon_{\mu \nu}
dx^{\mu}\wedge dx^{\nu},}
where $R$ is the two-dimensional scalar curvature and $\epsilon_{\mu
\nu}$
is the antisymmetric numeric tensor defined by $\epsilon_{12}=1$.
Since $s$ and the exterior differential $d$ on the two-dimensional
world-sheet anti-commute, the two-form in Eq. \euler\
gives rise
to the descent equations:
\eqn\descent{\eqalign{
s \sigma^{(2)} =& d \sigma^{(1)}\cr
s \sigma^{(1)} =& d \sigma^{(0)} \cr
s \sigma^{(0)} =& 0. \cr}}
The 0-form $\sigma^{(0)}$ and the 1-form $\sigma^{(1)}$
are computed to be
\eqn\observables{\eqalign{
\sigma^{(0)} =& {1\over 4\pi}\sqrt{g} \epsilon_{\mu \nu}
\left[{1\over 2}c^{\mu} c^{\nu}R + c^{\mu}D_{\rho}(\psi^{\nu\rho}
- g^{\nu\rho}\psi^{\sigma}_{\sigma}) + D^{\mu}\gamma^{\nu} -{1\over
4}\psi^{\mu}_{\rho}\psi^{\nu\rho}\right] \cr
\sigma^{(1)} =& {1\over 4\pi}\sqrt{g}\epsilon_{\mu\nu}\left[c^{\nu} R +
D_{\rho}(\psi^{\nu \rho} - g^{\nu\rho}\psi^{\sigma}_{\sigma})
\right]dx^{\mu}.\cr}}

$\sigma^{(0)}$ is a non-trivial class in the cohomology of $s$ acting on
the space of the {\it reparametrization-covariant} tensor fields.
One can verify explicitly that such cohomology is in one-to-one
correspondence with the {\it semirelative} state BRS cohomology
defined
on the state space of the theory, quantized on the infinite cylinder in
the conformal gauge \bci .

Since the superghosts $\gamma^{\mu}$ are commutative, one can build
an
infinite tower of cohomologically non-trivial operators by taking
arbitrary powers of $\sigma^{(0)}$:
\eqn\osser{\sigma_n^{(0)} \equiv (\sigma^{(0)})^n}
with $n=0,1,\ldots$ The corresponding 2-forms
\eqn\obtwon{\sigma_n^{(2)} = n (\sigma_n^{(0)})^{n-1} \sigma^{(2)}
+ {n(n-1)\over 2}(\sigma_n^{(0)})^{n-2}\sigma_n^{(1)}\wedge
\sigma_n^{(1)}}
all belong in the $s$-cohomology modulo $d$ on the space of the
reparametrization-covariant tensor fields.

In order to evaluate correlators of observables $\sigma_n$, the
choice of a Lagrangian is required. The theory being topological, the
choice of a Lagrangian amounts to fixing the gauge.

The gauge fixing choice is of course dependent on the gauge freedom,
which in turn depends on the particular correlation function being
computed.
Indeed, if all the involved observables correspond to the integral over
the Riemann surface of 2-forms such as $\sigma^{(2)}_n$, the gauge
freedom
corresponds to the whole set of supercoordinate transformations on the
Riemann surface. But if the observables involve the local operator
$\sigma^{(0)}_n(x)$ at some point $x$, we have to restrict the gauge
freedom to supercoordinate
transformations that leave this point fixed; correspondingly, the fields
$c$ and $\gamma$ must vanish at this point. This, of course,
transforms the Riemann surface into a punctured surface.
We must also remember that in the case of a
sphere the gauge group should not include the isometries of the sphere;
therefore the gauge supercoordinate transformations should leave three
distinguished points fixed; some or all of these points could possibly
coincide with the position of some local operator. An analogous remark
holds true for the torus.

In the following we shall limit our analysis to the case of local
operators, therefore the fields $c$ and
$\gamma$ vanish in a set of points including those where the observables
are sitting. In this situation the coordinates of the fixed points, with
the exclusion of three of them in the case of the sphere, and of one of
them for the torus, are among the moduli of the punctured surface.
One should also notice that, due to the local vanishing of $c$ and
$\gamma$, the 0-form $\sigma^{(0)}$ reduces to:
\eqn\effective{\sigma^{(0)}\rightarrow
{1\over 4\pi}\sqrt{g}\epsilon_{\mu\nu}\bigl( D^{\mu}\gamma^{\nu} -{1\over
4}\psi^{\mu}_{\rho}\psi^{\nu\rho}\bigr).}

Let $\mgn$ be the moduli space of two-dimensional Riemann surfaces
of a given genus $g$, and let $m=(m^i)$, with $i=1,\ldots, 6g-6+2n$,
be local coordinates on $\mgn$. Fixing the gauge means choosing
a background metric $\emunu (x;m)$ for each gauge
equivalence class of metrics corresponding to the point
$m$ of $\mgn$.

It is convenient to decompose $\emunu$ as follows:
\eqn\background{\emunu (x;m) \equiv \sqrt{\eta}
\ehatd(x;m) \equiv e^{{\bar\varphi}}
\ehatd (x;m),}
with ${\rm det}({\hat\eta})_{\mu\nu}=1$;
$\hatg_{\mu\nu}$ is given
by the analogous definition for $\gmunu$. We also introduce the
tensor density
$$\inoh \equiv \sqrt{g}(\psi^{\mu\nu} -\half
g^{\mu\nu}\psi^{\sigma}_{\sigma}),
$$
\noindent in correspondence with the traceless part of the gravitino
field.

$\emunu$ defines a gauge-slice on the field space whose associated
Lagrangian reads as follows \ref\torino{\TORINO},\ref\trieste{\TRIESTE},
\vv :
\eqn\lagra{
{\cal L} = s \left[\half b_{\mu \nu} (\ghat -\ehat) +
\half \beta_{\mu \nu}(\inoh - d_p\ehat ) + \chi \partial_{\mu}(\ghat
\partial_{\nu}(\varphi -{\bar \varphi}))\right].}
In Eq. \lagra\ we have introduced the ``exterior-derivative''
operator
$$d_p \equiv p^i {\partial \over \partial m^i},$$
\noindent where $p^i$ are the anticommuting supermoduli, with
$i=1,\ldots,6g-6+2n$, the superpartners of the commuting moduli $m^i$;
$\ab$, $\abeta$ and $\chi$
are the antighost fields, with ghost numbers $-1,-2$, and $0$
respectively. Their BRS transformation laws are given by
\eqn\multipliers{\eqalign{
&s\,\ab = \Lambda_{\mu \nu}\cr
&s\,\abeta = L_{\mu \nu}\cr
&s\,\chi = \lc \chi + \pi\cr }\qquad
\eqalign{&s\,\Lambda_{\mu \nu} = 0\cr
&s\,L_{\mu \nu} = 0 \cr
&s\,\pi = \lc \pi -\lgamma \chi,\cr}
}
where $\Lambda_{\mu\nu}$, $L_{\mu\nu}$ and $\pi$ are Lagrangian
multipliers.

At first sight it would seem that the Levi-Civita connection
in the covariant derivative acting on $\gamma$ in \effective, causes
$\sigma^{(0)}(x)$ to depend on the Liouville field $\varphi(x)$ defined
in Eq. \background.
However, the vanishing of $\gamma$ at point $x$ kills the
term in $\sigma^{(0)}$ dependent on the connection; therefore
the local observables $\sigma_n^{(0)}$
are actually invariant under Weyl transformations,
that is under $\varphi(x)$ translations.
Considering now the path integral that defines the correlation functions
and integrating out the Liouville
field and its superpartner (whose respective functional determinants
cancel by supersymmetry) one obtains a functional measure which
depends on the reduced, Weyl-invariant, background metric $\ehat
(x;m)$ but not on ${\bar\varphi}(x)$.

The moduli and the supermoduli are quantum mechanical variables,
which should be integrated over to obtain
the gauge-invariant correlators. It is therefore natural that
they transform under the action of the BRS operator $s$, according
to the following transformation laws \ref\bb{\BB},\torino:
\eqn\moduli{\eqalign{
s\, m^i = & \ci \cr
s\, p^i = & -\Gamma^i\cr}\qquad
\eqalign{s\,\ci =& 0 \cr s\,\Gamma^i = &0, \cr}}
where $\ci$ and $\Gamma^i$ are respectively anticommuting and
commuting Lagrange multipliers.

It is easy to check that the
action of $s$ on the background metric in the Lagrangian \lagra\
produces exactly the antighost insertions necessary
to gauge-fix the degeneracy associated to the zero-modes of
the antighost fields.

The Lagrangian in Eq. \lagra , written out in extended form, reads:
\eqn\lagratwo{\eqalign{
{\cal L} =&\, \half \Lambda_{\mu \nu}(\ghat -\ehat) + \half L_{\mu \nu}
(\inoh  - d_p\ehat )- \half \ab \lc \ghat  - \half \abeta \lgamma \ghat
\cr
& +\half \inoh \left[ (\lc\beta)_{\mu \nu} + \ab + 2\partial_{\mu}\chi
\partial_{\nu}(\varphi -{\bar \varphi})\right]\cr
&+\pi \partial_{\mu} (\ghat \partial_{\nu} (\varphi -{\bar
\varphi}))
- \chi \partial_{\mu} (\ghat \partial_{\nu}\psi^{\prime})\cr
&+\half\abeta d_{\Gamma}\ehat + \half\ab d_C \ehat +
\half\abeta d_p d_C \ehat + \chi \partial_{\mu}(\ghat
\partial_{\nu}d_C{\bar\varphi}),\cr}}
where
\eqn\psprime{\psi^{\prime} \equiv {\bar D}_{\sigma} c^{\sigma} +
\half\psi^{\sigma}_{\sigma},}
and where the notation $d_C\equiv \ci {\partial \over \partial m^i}$ and
$d_{\Gamma}\equiv \Gamma^i {\partial \over \partial m^i}$ has been
introduced.

Integrating out the Lagrange multipliers
$\Lambda_{\mu \nu}$, $L_{\mu\nu}$, $\pi$ and $\chi$
forces the metric and the gravitino field to take their background
values,
\eqn\substitutions{
\ghat \rightarrow \ehat ,\qquad
\varphi\rightarrow{\bar\varphi},\qquad
\inoh \rightarrow d_p \emunu, \qquad
\half\psi^{\sigma}_{\sigma}+ {\bar D}_{\sigma}c^{\sigma}\rightarrow
d_C{\bar\varphi},}
and the Lagrangian becomes
\eqn\newlagra{\eqalign{
{\cal L}^{\prime} = &\half\bigl[ -\ab \lc \ehat - \abeta \lgamma \ehat
+d_p\ehat (\lc \beta)_{\mu\nu}\cr
& +\ab (d_C \ehat - d_p \ehat ) + \abeta d_{\Gamma}\ehat +
\abeta d_p d_C \ehat \bigr] .\cr}}

In the following we will repeatedly make use of the fact that, when
the observables do not contain the antighost zero modes $b^{(i)}$,
integrating them out introduces into the correlators the factor
\eqn\simplela{\prod_i \delta (\ci - p^i).}

If moreover there are no antighost zero modes $\beta^{(i)}$ and no
antighost fields $\ab$ in the observables, one can integrate out
$\beta^{(i)}$ as well. This produces a further factor
\eqn\betaout{\prod_i \delta (\Gamma^i).}

We will consider correlators of the operator-valued 0-forms \osser\
obtained by functional averaging with respect to the
local quantum fields and the $\ci$ and $\Gamma^i$ multipliers,
collectively denoted by
$\Phi$, but not with respect to the moduli and supermoduli
$m^i$ and $p^i$:
\eqn\averages{\eqalign{
\Zmp \equiv &\Bigl\langle\prod_k
\sigma_{i_k}^{(0)}(P_{i_k})\Bigr\rangle
\cr
\equiv &\int\,[d\Phi] e^{-S(\Phi ;m^i,\,p^i)}
\prod_k \sigma_{i_k}^{(0)}(P_{i_k}).\cr}
}
The $0$-form $\sigma_{i_k}^{(0)}$ sits on the point $P_{i_k}$ of the
world-sheet manifold; therefore $\Zmp$ also depends on the choice of the
$P_{i_k}$'s. We are not considering this dependence explicitely since
it will disappear after moduli integration.

Because of ghost-number conservation, $\Zmp$ is a monomial of the
anticommuting supermoduli:
\eqn\ghost{\Zmp  = Z_{i_1\ldots i_N}(m^i)
p^{i_1}\ldots p^{i_N},}
where $N$ is the total ghost number of observables $\sigma_{i_k}^{(0)}$:
\eqn\ghnumber{N = \sum_k ({\rm ghost\#}\, \sigma_{i_k}^{(0)})
= 2 \sum_k i_k.}
Under a reparametrization ${\tilde m}^i = {\tilde m}^i(m)$ of the
local coordinates $m^i$ on the moduli space $\mgn$, the supermoduli
transform as:
$${\tilde p}^i = {\partial {\tilde m}^i \over \partial m^j}p^j.$$
We thus identify the anticommuting supermoduli with the
differentials on the moduli space, i.e. $p^i \rightarrow dm^i.$
Correspondingly, the function $\Zmp$ of the moduli and
supermoduli can be thought of as an {\it N}-form on the moduli space
$\mgn$, at least {\it locally} on $\mgn$. The question of whether or not
such a local form extends to a globally defined form on $\mgn$ will be
discussed in the next section.

Assume for the moment that the form  $\Zmp$ is globally defined on
$\mgn$. Whenever the following ghost number selection rule is
satisfied,
\eqn\selection{N = 2 \sum_k i_k = 6g-6 +2n,}
$\Zmp$ defines a measure on $\mgn$, which can be integrated
to produce some number. The collection of these numbers encodes some,
at least, of the gauge-invariant contents of 2D topological gravity.

It is easy to show that the action of the BRS operator $s$ on the
quantum
fields $\Phi$ translates into the action of the exterior differential
$d_p \equiv p^i\partial_i$ on the forms $\Zmp$. More precisely,
one can prove the following Slavnov-Taylor identities:
\eqn\identities{\eqalign{
(i)\;\;  &s\, O(\Phi^\prime) = 0 \Rightarrow d_p \langle O(\Phi^\prime)
\rangle = 0\cr
(ii)\;\; &O(\Phi^\prime) = s\, X(\Phi^\prime) \Rightarrow
\langle O(\Phi^\prime)\rangle  = d_p \langle X(\Phi^\prime)\rangle, \cr}}
where $O(\Phi^\prime)$ and $X(\Phi^\prime)$ are operators
that depend on the fields $\Phi^\prime$ other than $\ci$ and $\Gamma^i$
and do not contain the antighost zero modes and the antighost field $b$.

Let us prove, for example $(i)$. Denote by $\langle O\rangle^\prime$
the average with respect to the local fields $\Phi^\prime$ only,
without integrating over $\ci$ and $\Gamma^i$. Then
\eqn\closed{\eqalign{
\left[d_C - \Gamma^i{\partial\over\partial p^i}\right]\langle
O(\Phi^\prime)
\rangle^\prime =&\, \Bigl\langle \sum_{\Phi^\prime} s\,\Phi^\prime
{\delta S \over \delta
\Phi^\prime}O(\Phi^\prime)\Bigr\rangle^\prime  \cr
=&\, \bigl\langle s\, O(\Phi^\prime)\bigr\rangle^\prime \cr
=&\, 0.\cr}}
Since, under the state conditions, $\langle O\rangle^\prime$ is
proportional
to $\prod_i \delta(\ci -p^i)\delta(\Gamma^i)$, by integrating both sides
of  Eq. \closed\ with respect to the multipliers $\ci$ and $\Gamma^i$,
one obtains $(i)$. The proof of $(ii)$ is analogous.

\newsec{The main Ward identity}

The background metric $\emunu (x;m)$ cannot be chosen to be a everywhere
continuous function of $m$. In fact $\emunu (x;m)$ is a section of
the gauge bundle over $\mgn$ defined by the space of two-dimensional
metrics on a surface of given genus and $n$ punctures. This bundle
is non-trivial and therefore does not admit a global section.
It follows that $\emunu (x;m)$ must be a {\it local} section of
the bundle of two-dimensional metrics. Let $\{\Ua\}$ be a covering of
$\mgn$ of open neighbourhoods of $\mgn$, with $\bigcup_a \Ua =
\mgn$. The background gauge is defined by a collection
$\{\emunua (x;m)\}$ of two-dimensional metrics, with each $\emunua (x;m)$
locally defined, as a function of $m$, on $\Ua$.
The functional average in Eq. \averages\ defines
{\it local} closed forms $\Zmpa$, on each open set $\Ua$.

The collection of local forms $\{\Zmpa \}$  corresponds to
a {\it globally} defined form $\Zmp$ on $\mgn$ if and only if
\eqn\glo{\Zmpa = \Zmpb \qquad {\rm on}\quad \Uab \equiv \Ua
\cap \Ub .}
We will see that for general operators and background metrics,
$\{\Zmpa\}$ does not satisfy Eq. \glo , but rather an equation of the
more general form:
\eqn\local{\Zmpa -\Zmpb = d_p \Sab \qquad {\rm on}
\quad \Uab \equiv \Ua \cap \Ub .}
When $\Sab$ in Eq. \local\ is non-vanishing, $\{ \Zmpa\}$ does not
define a global form $\Zmp$, which could be integrated over $\mgn$ to
give a gauge-invariant expectation value. Under such circumstances,
one has to resort to the \v{C}ech notion of form cohomology
in order to obtain a gauge-invariant definition of
the ``integral'' of $\{ \Zmpa\}$ over $\mgn$. This will
be illustrated in the next section. Here, we will derive
a general formula for the $\Sab$ appearing in the Ward identity
\local .

Two background metrics $\emunua$ and $\emunub$ are related on the
intersection $\Uab$ by a combination of a diffeomorphism and
a Weyl transformation. One should remember however that
we  are considering in Eq. \averages\
expectation values of observables that involve only the 0-forms \effective\
which, as explained in the previous section, depend
on the reduced, Weyl-invariant, background metric $\ehat (x;m)$
but not on ${\bar\varphi}(x)$. Therefore we only need to consider
the Ward identities relative to diffeomorphisms of the background reduced
metric $\ehat (x;m)$.

Let therefore $O$ be the observable
\eqn\obserprod{O \equiv \prod_{i=1}^n \sigma^{(0)}_{n_i}(x_i)}
of total ghost number $\sum_i 2n_i = 2N$. The Ward identity that
we will prove in this section reads as:
\eqn\mainward{\bigl(\delta \langle O\rangle\bigr)_{ab} \equiv
\langle O\rangle_b - \langle O\rangle_a
= d_p \int_0^1 dt \int d[\Phi] {\rm e}^{- S_{ab}(t)}\bigl (I_b - I_a
\bigr)\, O (c, \gamma, \hatg, \hino ).}
Let us first explain the notation used in Eq. \mainward .
$\langle O\rangle_a$ and $\langle O\rangle_b$ denote the functional
averages of $O$ with backgrounds $\ehata$ and $\ehatb$ respectively.

$S_{ab}(t)$ is the following action depending on the real
parameter $t$
\eqn\interpoaction{ S_{ab}(t) \equiv \, S_0 -
s \int d^2\, x \abeta \bigl[ t\lebo
+ (1 - t)\leao\bigr],}
where we introduced a third background $\ehato$ and the corresponding
action $S_0$. $\ehato$ is related to the
backgrounds $\ehata$ and $\ehatb$ by two diffeomorphisms:
\eqn\diffeo{\xa \rightarrow \x0 (\xa;m) ,\qquad \xb \rightarrow \x0
(\xb;m),}
which may in general depend on $m$ and which
define the following forms $\va$ and $\vb$ on $\mgn$ with values
in the vector fields on the world-sheet:
\eqn\vector{\eqalign{
\va &\equiv p^i v^{\mu}_{i a} \equiv \partial_i x^{\mu}_0
(\xa;m)\vert_{\xa=\xa(\x0;m)}\cr
\vb & \equiv p^i v^{\mu}_{i b} \equiv \partial_i x^{\mu}_0
(\xb;m)\vert_{\xb=\xb(\x0;m)}.\cr}}
Finally $I_a$ and $I_b$ are operators which shift the ghost field
$\gamma^{\mu}$ by $\va$ and $\vb$, i.e.
\eqn\Idef{I_a \equiv -\int d^2\,x\, {\va}^{\mu}(x) {\delta\over\delta
\gamma^\mu}(x),}
and analogously for $I_b$.

Before plunging into the derivation of our Ward identity \mainward\ we
will specialize it to the case when $O = \sigma_1^{(0)}$. Since the
general observable has the product structure \obserprod, the
action of $I_a$ on it will be  expressible in terms of the action on the
operator $\sigma_1^{(0)}$. If we introduce the transition matrix
\eqn\transition{ (\Ma)_{\mu}^{\,\,\,\nu}
\equiv {\partial x_0^\nu\over \partial x^\mu_a} }
and define analogously the matrix $\Mb$, the Ward identity $\mainward$
becomes, for $O=\sigma_1^{(0)}$,
\eqn\wardspecial{
\eqalign{
(\delta \langle \sigma_1^{(0)}\rangle)_{ab} =&\, d_p\langle I_b
\sigma_1^{(0)}\rangle_0 - d_p \langle I_a \sigma_1^{(0)}\rangle_0 \cr
=&\, {1\over 4\pi} d_p \epsilon_{\mu\nu}{\hat\eta}^{\nu\lambda}
\bigl(\Mbi d_p\Mb-\Mai d_p\Ma \bigr)^{\,\,\,\mu}_\lambda .\cr }}
It is convenient to take the coordinates $x_0(x_a;m)$ to be
{\it isothermal} complex coordinates relative to the
complex structure $m$. That is, let $x_0(x_a;m) =
\bigl( Z_m(z_a,{\bar z}_a;m),{\bar Z}_m(z_a,{\bar z}_a;m)\bigr)$,
with $x_a = (z_a,{\bar z}_a)$ and
\eqn\isothermal{d\Zm \otimes d\Zmbar = |\If{a}|^2 (d\za +\bel{a}d\zabar)
\otimes (d\zabar + \bbel{a} d\za),}
where $\bel{a}$ and $\If{a}$ are corresponding Beltrami differentials
and integrating factors. Then, the transition matrix $\Ma$ becomes
\eqn\isotransition{ (\Ma)_{\mu}^{\,\,\,\nu} =
\left(\matrix{\If{a} & \bIf{a} \bbel{a} &\cr
\If{a} \bel{a} & \bIf{a} &\cr}\!\right),}
and the action of the operator $I_a$ on $\sigma_1^{(0)}$ reads
\eqn\isoshift{
f_a(x;m)\equiv \langle I_a \sigma_1^{(0)}(x)\rangle_0 =
{1\over 4\pi i} d_p\log {\If{a}\over \bIf{a}}
+ {1\over 4\pi i}{\bel{a}d_p\bbel{a} -\bbel{a}d_p\bel{a}\over
1-|\bel{a}|^2}.}
\bigskip
Eq. \mainward\ implies that $\{ \Zmpa \}$ is the restriction
of a globally defined form only for those backgrounds $\{ \het{a}\}$
for which $\va =\vb$.
This condition is equivalent to the requirement that $\{ \het{a} \}$
be a modular invariant global section on the space of reduced metrics
over Teichm\"uller space \torino,\trieste.
By this we mean that values of the section at points related by modular
transformations must differ by diffeomorphisms which are independent
of the Teichm\"uller point. In special cases -- for example,
on ${\cal M}_{1,1}$ -- sections $\{ \het{a}\}$
with this property can be found. Correspondingly, the local forms
$\{ \Zmpa\}$ associated to the correlator
$\langle\sigma_1\rangle_{g=0}$ match on the various patches of moduli
space to give a globally defined form \trieste.
However, for $\mgn$
with $g$ and $n$ generic, local sections with this property do not
exist. In fact the existence of a section with the stated property
would imply the existence of a homomorphism from the mapping class
group into the group of diffeomorphism.

Actually, the absence of modular invariant sections is fortunate since
the local expectation values $\langle O\rangle_a$ vanish
for almost all observables, as we will explain in detail in the
next section. Or, turning things around, the fact that correlators
of topological gravity are not trivial proves indirectly that
there is, generally, no section for which $\langle O\rangle_a$
glue together to produce a globally defined form.
One must therefore live with the general situation in which the
r.h.s. of Eq. \mainward\ does not vanish.
In the next section we will explain how to recover a global form
starting from \mainward\ and from a tower of descendant Ward identities.
\bigskip
PROOF OF EQ. \mainward . Let $\Phio(x)$, $\Phia(x)$ and $\Phib(x)$ denote
collectively the quantum fields in the coordinate systems $x_0$, $x_a$
and
$x_b$ respectively; $\Phio(x_0)$ and $\Phia(x_a)$ are related by a
diffeormorphism $x_0 = x_0(x_a;m)$, which generally depends on $m$. The
same diffeomorphism relates the background $\ehato$ and $\ehata$:
\eqn\metrictrans{\ehata (\xa;m)=
{1\over {\rm det}({\partial\xa\over \partial\x0})}
{\partial \xa^{\mu}\over \partial \x0^{\sigma}}
{\partial \xa^{\nu}\over \partial \x0^{\rho}}
{\hat\eta}_0^{\sigma\rho}(\x0;m).}
{}From Eq. \metrictrans\ one derives the transformation law for
the derivatives of the background with respect to the moduli:
\eqn\dmetric{
\partial_i{\hat\eta}^{\mu\nu}_a (\xa;m)\equiv
{\partial\, {\hat\eta}^{\mu\nu}_a\over \partial m^i}=
{1\over {\rm det}({\partial\xa\over \partial\x0})}
{\partial \xa^{\mu}\over \partial \x0^{\sigma}}
{\partial \xa^{\nu}\over \partial \x0^{\rho}}
\bigl( \partial_i {\hat\eta}^{\sigma\rho}_0 (\x0;m) +
({\cal L }_{v^i_a}{\hat\eta}_0)^{\sigma\rho}
(\x0;m)\bigr ). }
Let $S_0$ denote the action defined via the background metric
$\ehato$, that is:
\eqn\actiona{ \eqalign{S_0 &=\, s \int\! d^2\,\x0 \bigl[\ab (\ghat
-\ehato )(\x0) + \abeta
(\inoh - \dpeo )(\x0)\bigr] \cr
&= \int\! d^2\,\x0 \bigl[\Lambda_{\mu\nu}(\ghat -\ehato )(\x0)
+ L_{\mu\nu}(\inoh - \dpeo )(\x0)\cr
&\qquad\qquad - \ab\bigl(\inoh + (\lc\hatg)^{\mu\nu}-
d_{\cal C}\ehato\bigr)(\x0)\cr
&\qquad\qquad + \abeta \bigl((\lc\hino)^{\mu\nu}-
d_{\cal C}\dpeo + d_{\Gamma}\ehato \bigr))(\x0)\bigr] , \cr }}
where it should be recalled that
$$d_p = p^i{{\partial}\over {\partial m^i}}, \quad
d_{\cal{C}} = {\cal{C}}^i{{\partial}\over {\partial m^i}}, \quad
d_{\Gamma} = \Gamma^i{{\partial}\over {\partial m^i}}.$$
The crucial identity relating $S_0$ to $S_a$ is
\eqn\cruidentity{S_a (\Phia) = \Bigl[S_0 - s\,\int d^2x_0\,
\abeta \leao\Bigr]
(c_0-\vac ; \Lambda_0 - {\cal L}_{\vac}b_0; L_0 - {\cal L}_{\vac}
\beta_0; \Phi_0^\prime),}
where $\Phi_0^{\prime}$ denote the fields in the $x_0$ coordinate system
other than $c^{\mu}$, $\Lambda_{\mu\nu}$ and $L_{\mu\nu}$, and we defined
$\vac\equiv \ci v^i_a$.

The identity \cruidentity\ can be verified directly. A more conceptual
way to understand it is to recall that the BRS operator $s$ includes
an exterior derivative with respect to the moduli and that
the relation between the fields $\Phia$ and the fields $\Phio$
involves $m$-dependent transition functions. Therefore the $s$-variation
of a field in the $x_a$ reference frame does not transform
covariantly when expressed in terms of the fields in the $x_0$ reference
system. If one writes $s\Phia = {\hat\Phia}$ for the $s$-variation
of the fields in the $x_a$ reference system, one finds, by a calculation
identical to the one leading to Eq. \dmetric, that
\eqn\brstransition{s\,\Phio = {\hat\Phio} - {\cal L}_{\vac} \Phio,}
where ${\hat\Phio}$ is obtained from ${\hat\Phia}$ by means of a general
coordinate transformation.
To give an example, $s\, \ghat_0 = ({\cal L}_{c_0}\hatg_0)^{\mu\nu}
+ \inoh_0 - {\cal L}_{\vac}\hatg_0$.

Comparing Eq. \brstransition\ with the BRS transformation laws,
Eqs. \brs, \multipliers, one sees that the Lie derivative in the r.h.s.
of
Eq. \brstransition\ can be reabsorbed in the operator $s$ by means of the
following redefinition of the $c$ ghosts and the Lagrangian
multipliers :
\eqn\redefine{{\tilde c}^\mu = c_0^\mu - {\vac}^\mu ,\quad
{\tilde\Lambda}_{\mu\nu} = \Lambda_{\mu\nu} -
({\cal L}_{\vac}b_0)_{\mu\nu}, \quad
{\tilde L}_{\mu\nu} = L_{\mu\nu} - ({\cal L}_{\vac}\beta_0)_{\mu\nu}.}
For example, $s\, \ghat_0 = ({\cal L}_{c_0 -\vac}\hatg_0)^{\mu\nu}
+ \inoh_0$.

Thus, taking into account the transformation law of the background Eq.
\dmetric, one obtains for the action in the background $\het{a}$:
\eqn\cruderivation{
S_a = s \int d^2x_0 \bigl[(b_0)_{\mu\nu}\bigl(
\ghat_0 -\ehato\bigr)(x_0) + (\beta_0)_{\mu\nu}\bigl(
\inoh_0 -d_p\ehato - \leao\bigr)(x_0)\bigr],}
where the operator $s$ acts on the fields $\Phio$ according to
Eq. \brstransition. But since, after the redefinitions
in Eqs. \redefine, the action of $s$ on the fields $\Phio$
is the standard one, Eq. \cruderivation\ is equivalent
to the identity \cruidentity.

{}From now on, let us restrict ourselves to observables $O$ which are
reparametrization-invariant {\it and} independent from the
Lagrangian multipliers $\Lambda_{\mu\nu}$ and $L_{\mu\nu}$.
The functional average $\langle O\rangle_a$ in the background
$\het{a}$ can be rewritten in terms of the averages in the background
$\het{0}$ by means of the identity \cruidentity\ and after
a suitable change of integration variables:
\eqn\correlationa{
\langle O (c,\gamma,\hatg,\inoh)\rangle_a
=  \int d[\Phi]\,{\rm e}^{- S_0 + s\int d^2x\, \abeta\leao}
\,O(c+\vac,\gamma, \hatg,\inoh ).}
In a similar way, we derive
\eqn\correlationb{\langle O (c,\gamma,\hatg,\inoh)\rangle_b
=  \int d[\Phi]\, {\rm e}^{- S_0 + s\int d^2x\, \abeta\lebo}
\,O(c+\vbc,\gamma, \hatg,\inoh ).}

At this point we need to spell out a further crucial requirement
that must be satisfied by the observable $O$, that is
\eqn\equivariance{O(c + \vac, \gamma,\hatg,\inoh ) =
O (c,\gamma,\hatg,\inoh ).}
The coordinate of the points of the world-sheet where the local
observables are inserted do not change under the reparametrizations
\diffeo, i.e. $\vac (x_i)=0$, at the points $x_i$ where the
$\sigma_n^{(0)}(x)$
are considered. Therefore the condition \equivariance\ is satisfied
whenever the observables do not depend on {\it derivatives} of the
ghost $c$.  This is of course the case for the $\sigma_n^{(0)}$,
which are the observables studied in this paper. Eq. \equivariance\ is
the {\it equivariant} condition in the present context and is equivalent
to the $b_0^{-}$-condition of the operator formalism \bci.

Thus, assuming Eq. \equivariance , we arrive at
\eqn\proofone{\eqalign{
\langle O \rangle_b - \langle O \rangle_a =&
\int_0^1dt\, \int d[\Phi] {d\over dt}{\rm e}^{- S_{ab}(t)} O\cr
=& \int_0^1dt \, \int d[\Phi] \bigl[s \int d^2x\,\abeta
\bigl(\lebo -\leao\bigr)\bigr]{\rm e}^{-S_{ab}(t)} O\cr
=& \int_0^1dt \, \int d[\Phi]s \bigl[ \int d^2x\,\abeta
\bigl(\lebo -\leao\bigr){\rm e}^{-S_{ab}(t)} O\bigr],\cr}}
where $S_{ab}(t)$ is the interpolating action defined in Eq.
\interpoaction .

Notice that the functional integral of an $s$-variation
would vanish if only the action of $s$ on the local fields
(Eqs. \brs, \multipliers) were considered. However, since
$s$ acts also on moduli and supermoduli (Eq.\moduli\ ),
a non-vanishing result is obtained:
\eqn\prooftwo{\eqalign{
\bigl(\delta\langle O\rangle\bigr)_{ab}=
\int_0^1\! dt &\int \prod_i d{\cal{C}}^i\prod_j d\Gamma^j
\Bigl(d_{\cal C}- \Gamma^k{{\partial}\over {\partial p^k}}\Bigr)\cr
&\int d[\Phi^{\prime}]\int d^2x\,\abeta\bigl(\leao-\lebo\bigr)
{\rm e}^{- S_{ab}(t)}\, O,\cr}}
where $d[\Phi^\prime]$ denotes the functional integral with respect
to the fields other than $\ci$ and $\Gamma^i$. As stated before, $O$ does
not depend on $\Lambda_{\mu\nu}$ and thus one can safely substitute
$\ghat$ with $\ehat$ in the functional integral in the r.h.s. of
Eq. \prooftwo . Moreover, $I_a S_{ab}(t) = \int d^2x\, \abeta
(\lva\hatg)^{\mu\nu}$ and  $I_b S_{ab}(t) = \int d^2x\, \abeta
(\lvb\hatg)^{\mu\nu}$, where $I_a$ and $I_b$ are the operators defined
in Eq. \Idef. Therefore
\eqn\proofthree{ \eqalign{
\bigl(\delta\langle O\rangle\bigr)_{ab}
=& - \int_0^1\!\!dt\int\! \prod_i d\ci\prod_jd\Gamma^j
\Bigl(d_{\cal C}- \Gamma^k{\partial\over\partial p^k}\Bigr)
\int\! d[\Phi^\prime]\bigl[(I_a - I_a){\rm e}^{- S_{ab}(t)}\bigr]\, O\cr
=&\int_0^1\!\!dt\int\!\prod_i d\ci\prod_j d\Gamma^j
\Bigl(d_{\cal C}- \Gamma^k{\partial\over\partial p^k}\Bigr)
\int\! d[\Phi^\prime]{\rm e}^{- S_{ab}(t)}\bigl(I_b - I_a\bigr)\, O.\cr}}

Finally, since $O$ is independent of $\Lambda_{\mu\nu}$, $L_{\mu\nu}$,
$b_{\mu\nu}$ and
$\beta_{\mu\nu}$, the functional integral in the last formula is
proportional  to $\prod_i\delta (p^i-\ci)\delta(\Gamma^i)$
and hence one arrives at the Ward identity \mainward.

\newsec{\v{C}ech-De Rham cohomology and contact terms}

We have seen in the previous section that, for a generic background
gauge $\{ \het{a} \}$ with $m$-dependent transition functions
$\Ma (x;m)$,
the collection of local forms $\{ \Zmpa \}$ is not the restriction to
the cover $\{\Ua \}$ of a form globally defined on $\mgn$.
We want to show in this section that the Ward identities \mainward,
together with a whole chain of ``descendant'' Ward identities,
contain the necessary geometrical data to define a global
closed form on the moduli space.

Let $2N$ be the degree of the local forms $\Zmpa$, i.e. the ghost number
of the observable $O(c,\gamma,\hatg,\hino)$.
It is convenient to introduce formal anticommuting variables $\xi^a$
and to collect the local forms $\Zmpa$ into one single object:
\eqn\zeroco{Z^{(2N)}_0 = \sum_a \xi^a\Zmpa .}
$Z^{(2N)}_0$ is an element of $C^0(\mgn;\Omega^{(2N)})$, the space of
$0$-cochains with values in the local $2N$-forms
of $\mgn$. One defines analogously $C^q(\mgn;\Omega^{(p)})$ as
the space of $q$-cochains $Z^{(p)}_q$:
\eqn\qcochains{ Z^{(p)}_q = \sum_{a_0< a_1<\cdots
<a_q}Z^{(p)}_{a_0\ldots a_q}\xi^{a_0}\xi^{a_1}\ldots \xi^{a_q},}
with values in the local $p$-forms $Z^{(p)}_{a_0\ldots a_q}$ defined
on the $q$-fold intersections $\Uq \equiv
{\cal U}_{a_0}\,\cap\, {\cal U}_{a_1}\ldots\cap\,{\cal U}_{a_q}$.
(In Eq. \qcochains\ one is assuming that the set where
the indexes $a_0,a_1,\ldots ,a_q$ take values is ordered.)

By introducing the nilpotent coboundary operator
\eqn\coboundary{\delta \equiv \sum_a \xi^a ,}
one can recast the Ward identity \mainward\ in a compact form
\eqn\wardcompact{ \delta Z^{(2N)}_0 = d_p \Zf {1}.}
Since $d_p$ and the coboundary operator $\delta$ anticommute, Eq.
\wardcompact\ implies the existence of the familiar chain of descent
equations:
\eqn\chain {\eqalign
{\delta Z^{(2N)}_0 &= d_p \Zf 1  \cr
\delta \Zf 1 &= d_p \Zf 2  \cr
\cdots &= \cdots \cr
\delta Z^{(1)}_{2N-1} &= d_p Z^{(0)}_{2N} \cr
\delta Z^{(0)}_{2N} &= 0,\cr}}
where $\Zf q$ is a $q$-cochain with values in the space of local
$(2N-q)$-forms defined on the $q$-fold intersections
$\Uq$.

Let $\{Z^{(2N)}_0 , \Zf 1 , \ldots , Z^{(1)}_{2N-1} , Z^{(0)}_{2N}\}$
be a solution of the
chain of Ward identities in Eq. \chain ,  and consider the
following object:
\eqn\cdrcocycle{ Z_D =Z^{(2N)}_0 +\Zf 1 + \cdots + Z^{(0)}_{2N}. }
$Z_D$ is a {\it cocycle} of the \v{C}ech-De Rham complex
$C^*(\mgn; \Omega^*) \equiv \oplus_{p,q\ge 0} C^q (\mgn;\break
\Omega^p)$. This is the complex equipped with the nilpotent
operator $D \equiv d_p - \delta$. (Recall that the
variables $\xi^a$ anticommute by definition with the exterior
differential $d_p$.) In fact it is easily proved that
the Ward identities \chain\ are equivalent
to the single cocycle equation
\eqn\wardcdr{DZ_D =0.}
A well-known result of cohomology theory is that the cohomology of
the \v{C}ech-De Rham complex is isomorphic to the De Rham
cohomology of globally defined forms \ref\bott{\BOTT}.
This means that given a solution of
the Ward identities \chain, \wardcdr\ one can construct a globally
defined closed form on $\mgn$. The equivalence of \v{C}ech-De Rham
cohomology and De Rham cohomology rests on the existence of the
homotopy operator
\eqn\homotopy{ K\equiv \sum_a \rho^a (m) {\partial \over
\partial\xi^a},}
where $\rho^a(m)$ is a partition of unity on $\mgn$; $K$ has
the obvious but fundamental property
\eqn\kproperty{\{ K, \delta \}=1,}
so that it acts on cocycles as the inverse of $\delta$.
One can use $K$ to solve the last of the equations in \chain , and
write
\eqn\last{Z^{(0)}_{2N} = \delta K Z^{(0)}_{2N}.}
Plugging this expression into the equation before last in
\chain, one obtains:
\eqn\beforelast{\delta (Z^{(1)}_{2N-1} + d_p K Z^{(0)}_{2N})=0,}
which again can be solved in terms of the homotopy operator:
\eqn\solveminus{Z^{(1)}_{2N-1} = - d_p KZ^{(0)}_{2N} +
\delta K Z^{(1)}_{2N-1} + \delta K d_p KZ^{(0)}_{2N} .}
Continuing in this way, it is possible to climb the chain \chain\ up to
the top, and to derive:
\eqn\topchain{\delta \left( \sum_{q=0}^{q=2N} (d_p K)^q \Zf q \right)=
0.}
Equation \topchain\ implies that the components $\Zga$ of the following
$0$-cocycle
\eqn\formglobal{\Zglobal = \sum_a \xi^a\Zga =
 Z^{(2N)}_0 + d_p K \Zf 1 + \cdots + {(d_p K)}^{2N} Z^{(0)}_{2N}}
are the restrictions to $\Ua$ of a globally defined form $\Zg$ on
$\mgn$.

Suppose now that the degree $2N$ of the globally defined form $\Zg$
equals the dimension $6g-6+2n$ of $\mgn$. Then, $\Zg$ can be
integrated
over $\mgn$ and this integral can be directly expressed in terms
of the solutions of the Ward identities \chain . Let $\{\Cel{a}\}$
be a cell decomposition of $\mgn$, with $\Cel{a} \subset \Ua$, and let
$\Cq \equiv \cap_{i=1}^q\Cel{a_i}$ of codimension
$q$ in $\mgn$ oriented in such a way that the boundary of a cell
$\partial\Cel{a_0a_1\ldots a_q}$ satisfies:
\eqn\orient{\partial\Cel{a_0a_1\ldots a_q} =\cup_b\Cel{a_0a_1\ldots
a_qb},}
where we have introduced the convention that
$\Cel{a_0a_1\ldots a_q}$ is antisymmetric in its indices in the sense
that it changes orientation when
exchanging a pair of indices. We have defined
in this way
$q$-chains of cells of codimension $q$ that are adjoint to the
$q$-cochains defined above. Indeed, given a $q$-chain and a
$q$-cochain, we can define the integral:
\eqn\integralchain{\int_{{\cal C}_q}\Zf q\equiv \sum_{a_0 <a_1
\ldots<a_q} \int_{\Cq} \left (\Zf q \right)_{a_0a_1\ldots a_q}.}
Taking into account the anticommuting variables $\xi^a$ introduced
above, it is clear that this integral can be interpreted as a combined
ordinary and Berezin integral, that is:
\eqn\integralide{\int_{{\cal C}_q}=\sum_{a_0 <a_1
\ldots<a_q} \int_{\Cq}\ d\xi^{a_q} \ldots d\xi^{a_0} .}
Exploiting this formalism and the identity
\eqn\bereidentity{
\delta\int_{{\cal C}_q} + \quad (-1)^q\int_{{\cal C}_q}\delta =
{1\over q!}\sum_{a_0 a_1
\ldots a_q} \int_{\Cq}\ d\xi^{a_{q-1}} \ldots d\xi^{a_0},}
it is not difficult to prove that
\eqn\intchain{\int_{{\cal C}_q}d_p\Zf q
= (-1)^{q+1}\int_{{\cal C}_{q+1}}\!\delta\Zf q.}
Using Eqs. \formglobal\  and \intchain\ the integral of
the globally defined form $\Zg$ is expressed in terms of the integrals of
the cochains satisfying the Ward identities \chain :
\eqn\gintegral{ \int_{\mgn} \Zg = \sum_{q=0}^{q=2N} (-1)^q\sum_{a_0
<a_1 \ldots<a_q} \int_{\Cq} \bigl(\Zf q \bigr)_{a_0a_1\ldots a_q}.}

In the previous section we derived the following expression for $\Zf{1}$:
\eqn\onecocycle{ \bigl( \Zf{1} \bigr)_{ab} = \int_0^1 dt\int d[\Phi]
{\rm e}^{- S_{ab}(t)}\bigl(I_b - I_a\bigr) \, O.}
By repeated use of the main Ward identity \mainward\ one can derive
an analogous formula for the $\Zf{q}$.

Let us introduce the interpolating action
\eqn\interpolaction{ S_{a_0\ldots a_q}(t_0,\ldots,t_q)=
S_0 - s\,\sum_{k=0}^q t_k \int d^2x \abeta
\bigl(\Lv{a_k}\het{0}\bigr)^{\mu\nu}, }
depending on the $q+1$ real parameters $t_k$. Let us also use the
notation
\eqn\taverage{\eqalign{
\langle \langle O \rangle \rangle_{a_0,\ldots,a_q}
\equiv &\int_0^1 \prod_{k=0}^q dt_k \delta\Bigl(\sum_{k=0}^q t_k -1\Bigr)
\int d [\Phi] {\rm e}^{-S_{a_0\ldots a_q}(t_0,\ldots,t_q)}\, O \cr
= &\int_0^1 dt_0 \int_0^{1-t_0}\! dt_1 \ldots \int_0^{1-t_1 -\ldots
-t_{q-2}}\!\!\!\! dt_{q-1}\cr
&\qquad\int d [\Phi] {\rm e}^{-
S_{a_0\ldots a_q}(t_0,\ldots,t_{q-1}, 1-\sum_{k=0}^{q-1}t_k)} O,\cr}}
for the functional average of the observable $O$ with the interpolating
action \interpolaction\ together with the integration over the
parameters $t_k$ with the measure specified above.

In Appendix A, we prove the following formula for the components of
the cochain $\Zf {q}$:
\eqn\qcociclo{\left(\Zf {q}\right)_{a_0\ldots a_q} = \sum_{k=0}^q (-1)^k
\langle\langle \Ia{0}\ldots \vIa{k}\ldots\Ia{q} O
\rangle\rangle_{a_0\ldots a_q},}
where the check mark above $\Ia{k}$ means that this term should be
omitted.

It is clear from this expression that the last non-vanishing cochain in
the descent \chain\ is $\ZN$, since each of the operators $\Ia{k}$ eats
up one $\gamma$ field. From Eq. \qcociclo\ we have
\eqn\ncociclo{\eqalign{
\bigl(\ZN\bigr)_{a_0,\ldots,a_N} &= \sum_{l=0}^N (-1)^l\langle\langle
\Ia{0}\ldots\vIa{l}\ldots\Ia{N}O\rangle\rangle_{a_0,\ldots,a_N}. \cr
}}
Since $\Ia{0}\ldots\vIa{l}\ldots\Ia{N}O$ does not contain the gravitino
field but only the metric field, its functional average with the
interpolating
action \interpolaction\ is independent of the $t_k$ parameters. Moreover
\eqn\ncociclobis{
\int_0^1 \prod_{k=1}^N dt_k \delta\Bigl(\sum_{k=0}^N t_k -1\Bigr) =
{1\over N!}.}
Therefore
\eqn\ncociclotwo{ \ZN = \delta \SN, }
where $\SN$ is the $(N-1)$-cochain with values in the $N$-forms
\eqn\ncociclothree{\eqalign{
\SN &= {1\over
N!}\sum_{a_1<\ldots <a_N}\xi^{a_1}\ldots\xi^{a_N}
\langle\Ia{1}\ldots\Ia{N}\,O\rangle_0\cr
&= \Bigl({1\over
N!}\Bigr)^2\sum_{a_1,\ldots ,a_N}\xi^{a_1}\ldots\xi^{a_N}\langle\Ia{1}
\ldots\Ia{N}\,O\rangle_0\cr
&=\Bigl({1\over N!}\Bigr)^2(-1)^{{(N-1)N\over 2}}\langle I^N\,
O\rangle_0,\cr}}
and we introduced the operator $I \equiv \sum_a \xi^a I_a$.

Recalling the product structure \obserprod\ of the observables, one
finally obtains
\eqn\ncociclofour{\SN = {1\over N!}(-1)^{{(N-1)N\over 2}}
\prod_i f(x_i)^{n_i},}
where $f(x_i)$ is the $0$-cochain with values in the local $1$-forms of
$\mgn$ defined by Eq. \isoshift, that is
\eqn\zeroone{\eqalign{
f(x_i) \equiv &\sum_a \xi^a \langle I_a\sigma_1^{(0)}(x_i)\rangle_0 =
{1\over 4\pi}\sum_a \xi^a {\rm tr}\bigl(\epsilon\, \Mai d_p\Ma\bigr) \cr
=&{1\over 4\pi i}\sum_a\xi^a \Bigl[d_p\log {\If{a}\over \bIf{a}}
+ {\bel{a} d_p \bbel{a} -\bbel{a} d_p\bel{a}\over 1-|\bel{a}|^2}
\Bigr].\cr}}
\medskip
It should be noted at this point that the expectation
value of the superghost $\gamma$ vanishes in the functional measure
that we are considering (Eq. \lagratwo).
The reason is that the field $\beta$ conjugate to $\gamma$ appears only
in
the supercurrent $\lc\beta$. However, this latter term
is zero inside any correlator since it is linear in $c$ and no antighost
$b$ is present either in the measure or in the observables.

It follows that expectation values
$\langle O(c,\gamma ,\hatg,\hino)\rangle_a = \langle \prod_k
\sigma_{n_k}^{(0)}\rangle_a$ all
vanish, with the exception of those containing exclusively
dilatons $\sigma_1^{(0)}$ and puncture operators. (Such exceptional
correlators can only occur in geni 0 and 1.)
In fact, these exceptional correlators also vanish if one takes
a gauge for which $d_p\het{a}(x_i;m) = 0$ at the points $x_i$ where
the observables are inserted.

In other words the ``bulk'' term of the generic correlator -- i.e.
the top-component $Z^{(2N)}_0$ of the \v{C}ech-De Rham
cocycle $Z_D$ --  vanishes with the given action for 2D topological
gravity. However, our analysis makes it clear that this is not a
gauge-invariant statement: the ``bulk'' of the correlators is not
a globally defined form on moduli space. The higher order cochains
$\Zf{q}$ in Eq. \qcociclo\ solving the chain of Ward identities \chain\
should be included in the evaluation of integral \gintegral\ over moduli
space of the the gauge-invariant global form $\Zg$. It is worth
emphasizing that in our framework the ``contact'' terms --
i.e. the contribution of the higher-order cochains to the integrated
gauge-invariant correlators -- have been uniquely determined
by solving the local Ward identities \mainward, \chain\ characterizing
the functional measure of the theory.

The solution \qcociclo\ of the Ward identities \chain\
simplifies considerably in a gauge in which $d_p\het{a}(x_i;m) = 0$.
In this case both the top-component $Z^{(2N)}_0$
and all the descendant cochains $\Zf{q}$ vanish, with the exception
of $\ZN$. Therefore the integrated gauge-invariant correlators
\eqn\gintegralsimple{\eqalign{
\int_{\mgn} \Zg =&\, (-1)^N \sum_{a_0
<a_1 \ldots<a_N} \int_{{\cal C}_{a_0\ldots a_N}}
\bigl(\ZN \bigr)_{a_0a_1\ldots a_N}\cr
=&\, (-1)^{{N(N+1)\over 2}}{1\over N!}
\sum_{a_0<a_1\ldots <a_N} \int_{{\cal C}_{a_0\ldots a_N}}
\bigl(\delta \SN\bigr)_{a_0\ldots a_N},\cr}}
are rewritten, after taking into account Eq. \intchain:
\eqn\gintegralbis{\eqalign{
\int_{\mgn} \Zg = & (-1)^{{N(N-1)\over 2}}{1\over N!}
\sum_{a_1<\ldots<a_N} \int_{\partial{\cal C}_{a_1\ldots  a_N}}
\bigl(\SN\bigr)_{a_1\ldots a_N}.\cr}}

Moreoever $\SN$ in Eq. \gintegralbis\ is $d_p$-closed
since the $0$-cochains with values in the $1$-forms $f(x_i)$
are $d_p$-closed when  $d_p\het{a}(x_i;m) = 0$:
\eqn\zeroonesimple{ f(x_i) = {1\over 4\pi i}d_p \log {\If{a}\over \bIf{a}}
(x_i;m).}
Thus, only the cochains ${\cal C}_{a_1\ldots a_N}$
whose interiors contain points where some $\If{a_k}$ vanish
contribute to the sum \gintegralbis . The integrating factors $\If{a_k}$
have zeros at those points in moduli space which correspond to degenerate
Riemann surfaces. We conclude that, in this gauge, the computation
of globally defined correlators can be localized
at the boundary of moduli space of Riemann surfaces.

\newsec{Relation to the algebro-geometric formulation}

The observables of topological gravity are expected to correspond
to certain cohomology classes of ${\bar \mgn}$
(the compactification of $\mgn$ obtained adding curves with
double points) introduced by Morita \ref\morita{\MORITA},
Mumford \ref\mumford{\MUMFORD} and Miller \ref\miller{\MILLER}.
The intersection numbers of such classes have been computed by
Konsevitch \ref\konsevitch{\KONSEVITCH}. In this section we
will show that the \v{C}ech-De Rham cocycles \qcociclo\ we obtained
by solving the local Ward identities of 2D topological gravity
are indeed equivalent to the Mumford-Morita-Miller classes.

Consider the holomorphic line bundles
$\Li$ over ${\bar \mgn}$ whose fibers are the cotangent bundles at the
$n$ marked points $x_i$ of the Riemann surface. Let $c_1(\Li)$
be the first Chern class of $\Li$. The expectation \wittop\ is that
the global form corresponding to $\langle \sigma_n^{(0)}(x_i)\rangle$
be cohomologous to $c_1(\Li)^n$. Indeed, the \v{C}ech-De Rham
cocycle $\{ Z^{(2)}_0, Z^{(1)}_1, Z^{(0)}_2\}$ corresponding
to $\sigma_1^{(0)}(x_i)$ is, by virtue of our general formula
\qcociclo:
\eqn\fundcocycle{\eqalign{
\bigl(Z^{(2)}_0\bigr)_a =&\, {1\over 2\pi i} { d_p \bel{a}\wedge d_p
\bbel{a}\over \bigl( 1- |\bel{a}|^2\bigr)^2}\cr
\bigl(Z^{(1)}_1\bigr)_{ab} =&\, \langle I_b\sigma_1^{(0)}(x_i)\rangle_0 -
\langle I_a \sigma_1^{(0)}(x_i)\rangle_0 =\,
{1\over 4\pi i}
\Bigl( d_p\log {\If{b}\over \bIf{b}}- d_p\log {\If{a}\over \bIf{a}}\cr
+&\, {\bel{b} d_p \bbel{b} -\bbel{b} d_p\bel{b}\over 1-|\bel{b}|^2}
-{\bel{a} d_p \bbel{a} -\bbel{a} d_p\bel{a}\over 1-|\bel{a}|^2}\Bigr)\cr
\bigl(Z^{(0)}_2\bigr)_{abc} =&\, 0.\cr}}
The corresponding global $2$-form $Z^{(2)}$ can
 be integrated over any $2$-cycle $C^{(2)}$ of $\mgn$.
If $\Cel{a}$ is a cell decomposition of $C^{(2)}$ one has,
according to formula \gintegral:
\eqn\fundintegral{\eqalign{
\int_{C^{(2)}} Z^{(2)}= & \sum_a \int_{\Cel{a}}\bigl(Z_0^{(2)}\bigr)_a -
\sum_{a<b}\int_{\Cel{ab}}\bigl(Z_1^{(1)}\bigr)_{ab}\cr
= &\sum_a \int_{\Cel{a}}\! {1\over 2\pi i}{d_p \bel{a}\wedge d_p
\bbel{a}\over \bigl( 1- |\bel{a}|^2\bigr)^2}\cr
+ &\sum_a\int_{\partial\Cel{a}}\!{1\over 4\pi i}\left[
d_p\log {\If{a}\over\bIf{a}}+ {\bel{a}d_p\bbel{a}
-\bbel{a}d_p\bel{a}\over 1-|\bel{a}|^2}\right]\cr
= & \sum_a\int_{\partial\Cel{a}}{1\over 4\pi i}
d_p\log {\If{a}\over\bIf{a}}.\cr}}
The crucial assumption in Eq. \fundintegral\ was the smoothness
of the Beltrami differentials $\bel{a}(x_i;m)$
at all points in $C^{(2)}$, including the points corresponding
to surfaces with nodes. This justifies applying Stokes
theorem to cancel $\int_{\Cel{a}}{ d_p \bel{a}\wedge d_p
\bbel{a}\over \bigl( 1- |\bel{a}|^2\bigr)^2}$ with the
$\mu$-dependent part of $\int_{\partial\Cel{a}} Z^{(1)}_1$.
This regularity condition is natural from
the field theoretical point of view, since $\bel{a}$ is the
expectation value of the reduced metric field $\ghat$.
{}From the mathematical point of view, smoothness of the complex
structure at the marked points is a feature of the Mumford-Deligne
compactification ${\bar\mgn}$ of moduli space.

If $\mgn$ were smooth the integral in Eq. \fundintegral\ would equal
the number of zeros and poles of $\If{a}$ on $C^{(2)}$ taken with
opposite signs. Since $\mgn$ is an orbifold, this means that the
zeros and the poles might have  fractional weights corresponding
to the order of the orbifold singularity. At any rate the integral
in Eq. \fundintegral\ coincides with the integral
$\int_{C^{(2)}}c_1(\Li)$. For, a holomorphic section of
$\Li$ is given by $d\Zm = \If{a}( dz + \bel{a} d{\bar z})(x_i;m)$.
Since the zeros and poles of this holomorphic section are precisely
the zeros and poles of $\If{a}$,
$\int_{C^{(2)}}c_1(\Li)=\int_{C^{(2)}} Z^{(2)}$ for any $C^{(2)}$.

\newsec{$\langle \sigma_1\sigma_0^3\rangle_{g=0}$}

In this expression, and in the analogous ones describing the other
examples
we label by $\sigma_0$ the fixed points of the Riemann surface  that
are not associated with any operator. Hence
by $\langle \sigma_1 \sigma_0^3\rangle_{g=0}$ we mean the
vacuum expectation value of $\sigma_1$ on a sphere with four fixed points.
This is a 2-form on $\mfour$.
Let $x_1$ be the point on the Riemann sphere where $\sigma_1(x_1)$
is inserted, $m$ the associated complex modulus
and  $P_0=0$, $P_1=1$, and $P_2=\infty$ the points where the
$\sigma_0$ operators are located.

Consider a cell decomposition $\Cel{a}$, with $a=0,1,2,3$ of $\mfour$
where $\Cel{0}$, $\Cel{1}$ and $\Cel{2}$ are disks surrounding
$P_0$, $P_1$, and $P_2$ respectively, and $\Cel{3}$ is the closure of
the complement
in $\mfour$ of $\bigcup_{a=0,1,2} \Cel{a}$. In Appendix B, we
review the ``plumbing fixture'' construction and show explicitly
that one can choose $\bel{a}(x_1;m)= 0$ and $f_a (x_1) =
{1\over 4\pi i}\log{\If{a}\over \bIf{a}}(x_1;m)$ such that
\eqn\zerofactors{f_a(x_1)= d\theta_a,\quad {\rm for}\quad a=0,1,2,
\qquad f_3(x_1) =0,}
where $\theta_a \equiv \arg (m - P_a)$, for $a=0,1$ and
$\theta_2 \equiv \arg ({1\over m})$.

Taking into account the orientation of $\Cel{2}$ with respect to
$\Cel{0}$ and $\Cel{1}$, one obtains from Eq. \fundintegral\
\eqn\fourpuncture{\int_{\mfour}\langle \sigma_1 \sigma_0^3\rangle
 = 1+1 -1 = 1,}
in agreement with the dilaton equation.

\newsec{$\langle \sigma_1^2 \sigma_0^3\rangle_{g=0}$ and $\langle
\sigma_2 \sigma_0^4 \rangle_{g=0}$}

The non-vanishing correlators on ${\cal M}_{0,5}$ are
\eqn\mzerofive{ Z^{(4)}\equiv \langle\sigma_1^2\sigma_0^3\rangle
\quad {\rm and}\quad {\tilde Z}^{(4)}\equiv
\langle\sigma_2\sigma_0^4\rangle.}

${\cal M}_{0,5}$ is parametrized by two complex coordinates $m_1$ and
$m_2$ representing the positions on the complex sphere of the
two punctures $x_1$ and $x_2$.
Let us choose a gauge for which $\mu(x_i;m) =0$ for $i=1,2$. Thus,
as explained in section 4, the non-vanishing terms of the
\v{C}ech-De Rham towers are
\eqn\zzprime{ Z_2^{(2)} = {1\over 2}\delta\bigl(f(x_1) f(x_2)\bigr)
\quad {\rm and} \quad
{\tilde Z}_2^{(2)} = {1\over 2}\delta\bigl(f(x_1)^2\bigr).}

Let us denote by ${\cal N}_{1\alpha}$ the hypersurfaces of
${\cal M}_{0,5}$, of complex codimension 1 characterized
by the equation $m_1 = P_\alpha$, with $\alpha=0,1,2$ and $P_0=0$,
$P_1=1$ and $P_2 =\infty$. Analogously, let ${\cal N}_{2\alpha}$
be the hypersurface characterized by the equation
$m_2=P_\alpha$. Finally, let $\tilde{\cal N}$ be the hypersurface
with $m_1 =m_2$.

A generic point in ${\cal N}_{1\alpha}\cup{\cal N}_{2\alpha}\cup
\tilde{\cal N}$
represents  a $5$-punctured complex sphere with one node.
The nine points $m_{\alpha\beta}\in {\cal M}_{0,5}$
with $m_1 =P_\alpha$ and $m_2 = P_\beta$ where two or three
of these hypersurfaces intersect correspond
to $5$-punctured Riemann spheres with two nodes.
As explained in section 4, the computation of the correlators
can be localized around such points.

At the points $m_{\alpha\beta}$ with $\alpha\not= \beta$, two
hypersurfaces ${\cal N}_{1\alpha}$ and ${\cal N}_{2\beta}$ intersect.
Near each of these six points, there are
only two relevant cells of real codimension two, with torus
topology,
\eqn\doubleintersection{\eqalign{
{\cal C}_{0AB}=& \{ |m_1 -P_\alpha|
= \epsilon, \; |m_2-P_\beta |
=\epsilon^\prime\} \cr
{\cal C}^\prime_{0CD}=&\{ |m_1-P_\alpha| = \epsilon^\prime, \;
|m_2-P_\beta| =\epsilon \}}}
where $\epsilon < \epsilon^\prime$.

$\Cel{0AB}$ is the triple intersection of cells $\Cel{0}$,
$\Cel{A}$ and $\Cel{B}$  for which the $0$-cochain, with values
in the $1$-forms, $f_a(x_i) = {1\over 4\pi i} d_p \log{\If{a}
\over \bIf{a}}(x_i;m)$, can be chosen to be
\eqn\doublefactors{\eqalign{
& f_0(x_1)= f_0(x_2) =0 \cr
& f_A(x_1) = d\theta_1,\quad f_A(x_2) = d\theta_2 \cr
& f_B(x_1) = d\theta_1, \quad f_B(x_2)= 0,\cr}}
with $\theta_1 \equiv\arg( m_1 - P_\alpha)$ and
$\theta_2 \equiv\arg( m_2 - P_\beta)$. $\Cel{0CD}$ is the triple
intesection with $f_0(x_i)$, $f_C (x_i)$ and $f_D (x_i)$ defined
as in Eq. \doublefactors\ after exchanging $x_1$ and $x_2$.

{}From Eq. \zzprime\ one derives the components of $Z^{(2)}_2$
and ${\tilde Z}^{(2)}_2$ on $\Cel{0AB}$ and $\Cel{0CD}$:
\eqn\doublecomponents{\eqalign{
&\bigl(Z^{(2)}_2\bigr)_{0AB} = \bigl(Z^{(2)}_2\bigr)_{0CD}
= \half d\theta_1 \wedge d\theta_2 \cr
&\bigl({\tilde Z}^{(2)}\bigr)_{0AB} =
\bigl({\tilde Z}^{(2)}\bigr)_{0CD} = 0.\cr}}

Near each of the points $m_{\alpha\alpha}$ there is,
beyond two cells analogous
to $\Cel{0AB}$ and $\Cel{0CD}$, a third relevant cell,
\eqn\tripleintersection{ \Cel{0FD} = \{ |m_1 -m_2| = \epsilon,
\quad |m_2 - P_\alpha| = \epsilon^\prime \},}
which is the triple intersection of cells $\Cel{0}$ $\Cel{F}$
and $\Cel{E}$, where
\eqn\triplefactors{\eqalign{
& f_0(x_1)= f_0(x_2) =0 \cr
& f_E(x_1) = d\theta_1,\quad f_E(x_2) = d\theta_2 \cr
& f_F(x_1) = f_F(x_2)= d \theta_3,\cr}}
where $\theta_3 \equiv \arg (m_1 -m_2)$. From Eq. \triplefactors\ one
derives the components of $Z^{(2)}_2$ and ${\tilde Z}^{(2)}_2$
on $\Cel{0EF}$:
\eqn\triplecomponents{\eqalign{
\bigl(Z^{(2)}_2\bigr)_{0EF} = &\half \bigl( f_E(x_1) f_F(x_2) +
f_F(x_1) f_E(x_2)\bigr)= \half \bigl(d\theta_1 +d\theta_2\bigr)
\wedge d\theta_3\cr
=& d\theta_2 \wedge d\theta_3 \cr
\bigl({\tilde Z}^{(2)}\bigr)_{0EF} = &\half\bigl( 2 f_E(x_1) f_F(x_1)
\bigr) = d\theta_2 \wedge d\theta_3 ,\cr}}
where we took into account that $\bigl(d\theta_1 - d\theta_2\bigr)
\wedge d\theta_3 = 0$ on $\Cel{0EF}$.

To sum the contributions of the various cells to the integrated
correlator, we need only to take into account the relative
orientations of the triple intersections at each of the nine
points $m_{\alpha\beta}$.

Let us first consider $Z^{(4)} =\langle \sigma_1^2 \sigma_0^3\rangle$.
The union of cells of type $\Cel{0AB}\cup\Cel{0CD}$ give $\pm 1$ for each
of the nine points $m_{\alpha\beta}$. It is straightforward to
verify that the contribution is $1$ for the five
points $m_{\alpha\beta}$ with
$\alpha, \beta = 0,1$ and $\alpha=\beta=2$, and $-1$ for the other
four points. The total is $1$. The cells of type $\Cel{0EF}$ give
instead $1$ for $m_{\alpha\alpha}$ with $\alpha =0,1$ and
$-1$ for $\alpha =2$, for a total of $1$. Thus,
\eqn\twodilaton{ \int_{{\cal M}_{0,5}}
\langle \sigma_1^2 \sigma_0^3\rangle = 2,}
in agreement with the dilaton equation.

Only cells of type $\Cel{0EF}$ contribute instead to ${\tilde Z}^{(4)}
=\langle \sigma_2 \sigma_0^4\rangle$, the ones near $m_{00}$ and
$m_{11}$ giving $1$ each and the cell near $m_{22}$ giving $-1$. Hence,
\eqn\fourpuncture{ \int_{{\cal M}_{0,5}}
\langle\sigma_2 \sigma_0^4\rangle = 1,}
as predicted by the puncture equation.

\newsec{Conclusions}

We have elucidated an intriguing question arising in the context of
2D topological gravity. The Lagrangian that leads
to a free superconformal model
also leads to a functional measure, for which averages of all equivariant
observables vanish, locally on the moduli space. The non-trivial
content of the theory should be encoded in contact terms sitting at
the boundary of moduli space, but it was not easy to see, in
the usual framework, how to compute these contact terms
directly from the functional integral.
We showed that the local Ward identities
characterizing the dependence of the functional measure on the
background gauge do capture the contact terms at the infinity in moduli
space.
Indeed the distinction between ``bulk'' and ``contacts'' is not
a gauge-invariant one. Contacts are required precisely for restoring
gauge invariance of the (possibly vanishing) ``bulk'' part of the
correlators.
We have shown that the ``bulk'' of correlators
of equivariant observables is not globally defined because
of the non-trivial dependence of the observables on the superghost.

We think that our analysis, beyond providing an intellectually satisfying
understanding
of contact terms in the particular context of 2D topological
gravity,  has a general validity and therefore might prove useful in
situations where
contacts have not yet been computed. We are thinking for example
of the exotic type topological string models relevant for 2D QCD \cmr\ or
of superstring models. We also expect that our approach should
clarify the nature of the holomorphic anomaly discovered in
the context of topological gravity coupled to $N=2$ superconformal
models \bcov.

Finally, and more speculatively, the mechanism to implement global BRS
invariance that we have analysed in 2D topological gravity might be of
some
relevance to analogous global non-perturbative issues associated with
the Gribov horizon in 4D non-Abelian gauge theories.
\bigskip
{\bf Acknowledgements}

It is a pleasure to thank R. Collina and R. Stora for many useful
discussions.
We also would like to thank E. Verlinde for interesting comments
and for bringing to our attention
Ref. \ref\hv{\HV},
in which the relevance of \v{C}ech-De Rham cohomology in the
context of superstring had been pointed out.
This work is partially supported by the ECPR, contract SC1-CT92-0789.

\appendix{A}{Proof of Eq. \qcociclo}

In this appendix we prove formula \qcociclo\
for the components of  the  cochain $\Zf {q}$:
\eqn\qcociclo{\left(\Zf {q}\right)_{a_0\ldots a_q} = \sum_{k=0}^q (-1)^k
\langle\langle \Ia{0}\ldots \vIa{k}\ldots\Ia{q} O
\rangle\rangle_{a_0\ldots a_q},}
where the check mark above $\Ia{k}$ means that this term should be
omitted.

We need to show that $\delta \Zf {q} = d_p \Zqplus $.  One has
\eqn\dimone{\eqalign{
\bigl(\delta \Zf{q}\bigr)_{a_0\ldots a_{q+1}} =&
\sum_{l=0}^{q+1} (-1)^l \bigl(\Zf{q}\bigr)_{a_0\ldots \check{a}_l\ldots
a_{q+1}}\cr
=&\sum_{l=0}^{q+1} (-1)^l \langle\langle\bigl[ \sum_{k=0}^{l-1}(-1)^k
\Ia{0}\ldots\vIa{k}\ldots\vIa{l}\ldots\Ia{q+1}\cr
&\quad -\sum_{k=l+1}^{q+1}(-1)^k
\Ia{0}\ldots \vIa{l}\ldots\vIa{k}\ldots \Ia{q+1}\bigr] O
\rangle\rangle_{a_0\ldots \check{a}_l\ldots a_{q+1}}\cr
=& \sum_{k<l}(-1)^{l+k} \big[ \langle\langle
\Ia{0}\ldots \vIa{k}\ldots\vIa{l}\ldots
\Ia{q+1}O\rangle \rangle_{a_0\ldots \check{a}_l\ldots a_{q+1}}\cr
& -\langle\langle\Ia{0}\ldots \vIa{k}\ldots\vIa{l}\ldots \Ia{q+1}O
\rangle\rangle_{a_0\ldots \check{a}_k\ldots a_{q+1}}\bigr].\cr}
}
According to the Eq. \taverage,
the functional measure associated with the right-hand side of \dimone,
that is the measure corresponding to $\langle\langle
\ldots\rangle \rangle_{a_0\ldots \check{a}_l\ldots a_{q+1}}
-\langle\langle\ldots\rangle\rangle_{a_0\ldots \check{a}_k
\ldots a_{q+1}}$,
for a suitable choice of the  $t$ variables, can be written in the form
\eqn\difmeas{\eqalign{
&\int_0^1\! \prod_{i=0,\, i\not= k,l}^{q+1} dt_i
\, dt' \delta(t'+\sum_{i=0,\, i\not= k,l}^{q+1} t_i -1)
 \Bigl({\rm e}^{-S_{a_0\ldots a_k\ldots\check{a}_l,\ldots, a_{q+1}}
(t_0,\ldots,t', \ldots,\check{t}_l,\ldots,t_{q+1})}\cr &\quad\quad
- \,{\rm e}^{-S_{a_0\ldots\check{a}_k\ldots a_l\ldots,a_{q+1}}
(t_0,\ldots ,\check{t}_k,\ldots,t',\ldots,t_{q+1})}\Bigr).\cr}}
Taking into account that
\eqn\pardev{{\partial\over\partial t_k}
S_{a_0\ldots a_k\ldots a_{q+1}}
(t_0,\ldots,t_k,\ldots,t_{q+1})=  - s \int d^2\, x\abeta
\bigl(\Lv{a_k}{\hat \eta}_0\bigr)^{\mu \nu},}
we transform \difmeas\ into the interpolating measure:
\eqn\intmeas{\eqalign{
&\int_0^1\!\prod_{i=0,\, i\not= k,l}^{q+1}\!\! dt_i
\, dt' \delta\Bigl(t'+\sum_{i=0,\, i\not= k,l}^{q+1} t_i - 1 \Bigr)
\int_0^1 du\, t'\cr
&\Bigl[ s \int d^2x\abeta\bigl(\Lv{a_k}\het{0} -
\Lv{a_l}\het{0}\bigr)^{\mu\nu}\Bigr]
{\rm e}^{-S_{a_0\ldots a_{q+1}}
(t_0,\ldots ,(1-u)t',\ldots,ut',\ldots ,t_{q+1})}\, ,\cr}}
which, after the relabelling $(1-u)\,t'\rightarrow t_k$ and
$u\,t'\rightarrow t_l$, becomes
\eqn\intmea{\int_0^1\!\prod_{i=0}^{q+1} dt_i \delta
\Bigl(\sum_{i=0}^{q+1} t_i - 1\Bigr)
\Bigl[ s \int d^2x\abeta\bigl(\Lv{a_k}\het{0} -
\Lv{a_l}\het{0}\bigr)^{\mu\nu}\Bigr]
{\rm e}^{-S_{a_0\ldots a_{q+1}}(t_0,\ldots ,t_{q+1})}\, .}
Therefore we can rewrite \dimone\ as follows:
\eqn\dimthree{\eqalign{
\bigl(\delta \Zf{q}\bigr)_{a_0\ldots a_{q+1}}\!\! & =
\sum_{k<l}(-1)^{l+k}\Bigl\langle\Bigl\langle \Bigl[ s \int d^2x
\abeta\bigl(\Lv{a_k}\het{0} - \Lv{a_l}\het{0}\bigr)^{\mu\nu}\Bigr]\cr
&\quad\Ia{0}\ldots \vIa{k}\ldots\vIa{l}\ldots
\Ia{q+1}O\Bigr\rangle\Bigr\rangle_{a_0\ldots a_{q+1}}\cr
&=\sum_{k<l}(-1)^{l+k}d_p \Bigl\langle\Bigl\langle\int d^2x\,
\abeta\bigl(\Lv{a_k}\het{0} - \Lv{a_l}\het{0}\bigr)^{\mu\nu}
\Ia{0}\ldots\vIa{k}\cr
&\quad\ldots\vIa{l}\ldots
\Ia{q+1}O\Bigr\rangle\Bigr\rangle_{a_0\ldots a_{q+1}}\cr
&+\sum_{k<l}(-1)^{l+k}\langle\langle
\bigl[\bigl(\Ia{k}-\Ia{l}\bigr)S_{a_0\ldots a_{q+1}}\bigr]\, s\,
\Ia{0}\ldots \vIa{k}\cr
&\quad\ldots\vIa{l}\ldots
\Ia{q+1}O\rangle\rangle_{a_0\ldots a_{q+1}}\cr}}
where we first made use of the Slavnov-Taylor identity
\identities\ $(ii)$, then substituted
$\ghat$ with $\ehat$ and finally took into account the equations
\eqn\dimthreebis{\Ia{k,l} S_{a_0\ldots  a_{q+1}}= \int d^2x\,
\abeta\bigl(\Lv{a_{k,l}}\hatg\bigr)^{\mu\nu}.}
Integrating by parts the operators $\Ia{k}$ and $\Ia{l}$ in \dimthree\
we obtain
\eqn\dimthreetris{\eqalign{
\bigl(\delta \Zf{q}\bigr)_{a_0\ldots a_{q+1}}\!\!\!=&
\sum_{k<l}(-1)^{l+k}d_p \langle\langle \bigl(\Ia{k} -\Ia{l}\bigr)
\Ia{0}\ldots \vIa{k}\cr
&\quad\ldots\vIa{l}\ldots\Ia{q+1}O\rangle\rangle_{a_0\ldots a_{q+1}}\cr
+&\sum_{k<l}(-1)^{l+k}\langle\langle (\Ia{k} -\Ia{l})\,s\,\Ia{0}\ldots
\vIa{k}
\cr &\quad\ldots\vIa{l}\ldots\Ia{q+1}O\rangle\rangle_{a_0\ldots a_{q+1}}
\cr = & \sum_{k<l}(-1)^{l+k}\langle\langle
(\{s,\Ia{k}\} -\{s,\Ia{l}\})\Ia{0}\ldots\vIa{k}\cr
&\quad\ldots\vIa{l}\ldots
\Ia{q+1}O\rangle\rangle_{a_0\ldots a_{q+1}}.\cr}
}
Now, $\{ s,\Ia \, \}$ is easily computed to be:
\eqn\antic{\eqalign{
&\int d^2\,x\,\bigl[\bigl(
\Gamma^iv_{ia}^{\mu}(x) - d_C {\va}^{\mu}(x)  +{\cal L}_c {\va}^{\mu}(x)
\bigr) {\delta\over\delta
\gamma^\mu}(x)\cr & - {\va}^{\mu}(x) {\delta\over\delta
c^\mu}(x) + {\cal L}_{{\hat v}_a} g^{\mu\nu}(x){\delta\over\delta
\psi^{\mu\nu}}(x)\bigr].\cr}}
which is a functional partial differential operator that commutes with any
$ I_b $. Hence, we obtain from \dimthreetris\ the desired result:
\eqn\dimfour{\eqalign{
\bigl(\delta \Zf{q}\bigr)_{a_0\ldots a_{q+1}}\!\!\!\! &
=\,\sum_l (-1)^{l}\sum_{k=0}^{l-1} (-1)^k \langle\langle
\{s,\Ia{k}\}\Ia{0}\ldots \vIa{k}\cr
&\quad\ldots \vIa{l}\ldots\Ia{q+1}O\rangle\rangle_{a_0\ldots a_{q+1}} \cr
&+\,\sum_l (-1)^{l}\!\sum_{k=l+1}^{q+1} (-1)^{k+1} \langle\langle
\{s,\Ia{k}\}\Ia{0}\ldots \vIa{l}\cr
&\quad\ldots \vIa{k}\ldots\Ia{q+1} O\rangle\rangle_{a_0\ldots a_{q+1}} \cr
& =\, \sum_l (-1)^{l}\langle\langle s\,\bigl(
\Ia{0}\ldots\vIa{l}\ldots\Ia{q+1}
O\bigr)\rangle\rangle_{a_0\ldots a_{q+1}} \cr
& =\,
d_p \sum_l (-1)^{l}\langle\langle\Ia{0}\ldots\vIa{l}\ldots\Ia{q+1}
O \rangle\rangle_{a_0\ldots a_{q+1}},\cr}}
where, in the second term in the right-hand-side of the first identity,
we have exchanged $k$ with $l$ and the order of the two sums.

\appendix{B}{The plumbing fixture}

We consider a cover of $\mfour$ consisting of three open disks
$\Ua$, with $a=0,1,2$, centred around the three
punctures $P_a$. The radius of these disks will be determined
in the following. A cell decomposition is obtained from this cover by
taking $\Cel{a}$ with $a=0,1,2$ to be proper closed subsets of
$\Ua$, and $\Cel{3}$ to be the closure of the complement of
$\Cel{0}\cup\Cel{1}\cup\Cel{2}$.

On the open patches $\Ua$ we consider maps
$\Map{a}: (\za,\zabar) \mapsto (\Zm, \Zmbar)$

\eqn\amaps{ \Map{a} = h_a^{-1}\circ F_{q_a(m)}\circ h_a,}
which carry from the coordinates $\xa$ to isothermal coordinates
\isothermal; $h_a(\za)$ are conformal maps of the complex sphere, and
$F_q(z,{\bar z})$ is defined by
\eqn\plufix{F_{q}(z,{\bar z}) \equiv z q^{\theta_R (1-|z|^2)}.}
$\theta_R(x)$ is a regularized  step function,
interpolating smoothly between $0$ and $1$:
\eqn\stfun{\theta_R(x) =\cases{0, & if $x\le 0$;\cr 1, &if $x\ge 1-
C$,\cr}}
with $0<C<1$. $F_q$ is a smoothened version of the quasi-conformal
map which defines the plumbing-fixture construction.

The plumbing-fixture parameters $q_a(m)$ are determined in terms of
the holomorphic $h_a$:
\eqn\qpar{q_a(m) \equiv {h_a(m)\over h_a(x_1)},}
where $x_1$ is the position where the $\sigma_1$ operator is inserted.
$h_a$ are chosen to be:
\eqn\holoa{h_a(z)=\cases{z, &if $a=0$;\cr z-1 , &if $a=1$;\cr
{1\over 2z-1},&if $a=2$.\cr}}
It is necessary that
\eqn\pcond{\Map{a}(P_a) = P_a\quad {\rm and}\quad
\Map{a}(x_1) = m}
for the maps $\Map{a}$ to define isothermal
coordinates corresponding to the complex structure $m$.
It is straightforward
to verify that Eqs. \pcond\ are satisfied as long as
\eqn\psati{{\rm max}\Bigl\{|x_1|^2, |x_1-1|^2, {1\over |2x_1-1|^2}\Bigr\}
\leq C.}
We can meet this condition by choosing, for concreteness, $x_1=1/2 +
i\sqrt{2}/2$ and $C=3/4$.

The open sets $\Ua$ of $\mfour$ on which the maps $\Map{a}$ are
invertible are defined by the equations
\eqn\opsets{0  < |q_a(m)| < R_0 \equiv \exp
\left[ {1 \over \sup_{x \ge 0}\left( 2x\theta_R (1-x)\right)}\right].}
It is important that $R_0>1$ for $\cup_a \Ua$ to cover the whole
$\mfour$. It is not difficult to check that in our concrete example
we can take $R_0 = 2/\sqrt{3}$. Then, the three open sets
which cover $\mfour$ are explicitly
\eqn\explicitsets{
\UU{0} = \{m:\,|m| <1 \},\quad \UU{1}=\{m:\, |m-1| <1\},\quad
\UU{2}=\bigl\{m:\, |m-1/2| > {\sqrt{3}\over 2\sqrt{2}}\bigr\}.}
The Beltrami differentials $\bel{a}(\za,\zabar;m)=
{\partial \Map{a}\over\partial \zabar}/{\partial \Map{a}\over\partial
\za}$ vanish identically at the point $x_1$:
\eqn\vanbel{\bel{a}(x_1;m) =0, \qquad \forall m\in \Ua .}
The integrating factors $\If{a}(\za,\zabar;m)={\partial \Map{a}\over
\partial \za}$ evaluated at the point $x_1$ are
\eqn\factp{\If{a}(x_1;m) = {(\log h_a)^{\prime}_{z=x_1}\over
(\log h_a)^{\prime}_{z=m}} = \cases{ {m\over x_1} & for $a=0$;\cr
{m-1\over x_1-1} & for $a=1$;\cr
{m-\half\over x_1-\half} & for $a=2$.\cr}}
Thus, by taking $\Cel{a}\subset \Ua$ for $a=0,1$ and $2$, one obtains
$f_a (x_1) = d\theta_a$, as stated in Eq. \zerofactors. On the
complement of $\Cel{0}\cup\Cel{1}\cup\Cel{2}$, $\If{a}(x_1;m)$
has neither zeros nor poles, and hence we can
equivalently set $f_3(x_1) =0$.

\listrefs

\bye